\documentclass[11pt]{article}
\pdfoutput=1
\usepackage{fullpage}
\usepackage{xcolor}
\usepackage{cite}
\usepackage{graphicx}
\usepackage{tensor}
\usepackage{amsmath}
\usepackage{amssymb}
\usepackage{subcaption}
\usepackage[utf8x]{inputenc} 
\title{}
\date{}
\renewcommand{\vec}[1]{\mbox{\boldmath$ #1 $}}

\newcommand{\Ad}{\dot{A}}
\newcommand{\Bd}{\dot{B}}
\newcommand{\Cd}{\dot{C}}
\newcommand{\Dd}{\dot{D}}

\def\beq{\begin{equation}}
\def\eeq{\end{equation}}
\allowdisplaybreaks
\begin{document}
\bibliographystyle{utphys}

% Commands
\newcommand{\be}{\begin{equation}}
\newcommand{\ee}{\end{equation}}
\newcommand\n[1]{\textcolor{red}{(#1)}} %in-text notes
\newcommand{\diff}{\mathop{}\!\mathrm{d}}
\newcommand{\lb}{\left}
\newcommand{\rb}{\right}
\newcommand{\f}{\frac}
\newcommand{\pd}{\partial}
\newcommand{\tr}{\text{tr}}
\newcommand{\fdiff}{\mathcal{D}}
\newcommand{\im}{\text{im}}
\let\caron\v
\renewcommand{\v}{\mathbf}
\newcommand{\T}{\tensor}
\newcommand{\R}{\mathbb{R}}
\newcommand{\C}{\mathbb{C}}
\newcommand{\Z}{\mathbb{Z}}
\newcommand{\msbar}{\ensuremath{\overline{\text{MS}}}}
\newcommand{\DIS}{\ensuremath{\text{DIS}}}
\newcommand{\abar}{\ensuremath{\bar{\alpha}_S}}
\newcommand{\bb}{\ensuremath{\bar{\beta}_0}}
\newcommand{\rc}{\ensuremath{r_{\text{cut}}}}
\newcommand{\Nd}{\ensuremath{N_{\text{d.o.f.}}}}
\newcommand{\red}[1]{{\color{red} #1}}
%\setlength{\parindent}{0pt}
% Nathan's macros
\newcommand{\mf}[1]{\mathfrak{#1}}
\newcommand{\cl}[1]{\mathcal{#1}}
\renewcommand{\[}{\begin{equation}\begin{aligned}}
\renewcommand{\]}{\end{aligned}\end{equation}}
\renewcommand{\Re}{\operatorname{Re}}
\renewcommand{\Im}{\operatorname{Im}}
\titlepage
%\begin{flushright}
%QMUL-PH-22-??\\
%\end{flushright}

\vspace*{0.5cm}

\begin{center}
{\bf \Large Deriving Weyl double copies with sources}

\vspace*{1cm} 
\textsc{Kymani Armstrong-Williams$^a$\footnote{k.t.k.armstrong-williams@qmul.ac.uk}, Nathan Moynihan$^a$\footnote{n.moynihan@qmul.ac.uk}
and Chris D. White$^a$\footnote{christopher.white@qmul.ac.uk}} \\

\vspace*{0.5cm} $^a$ Centre for Theoretical Physics, Department of
Physics and Astronomy, \\
Queen Mary University of London, 327 Mile End
Road, London E1 4NS, UK\\

\end{center}

\vspace*{0.5cm}

\begin{abstract}
  The Weyl double copy is a relationship between classical solutions
  in gauge and gravity theories, and has previously been applied to
  vacuum solutions in both General Relativity and its
  generalisations. There have also been suggestions that the Weyl
  double copy should extend to solutions with non-trivial sources. In
  this paper, we provide a systematic derivation of sourced Weyl
  double copy formulae, using spinorial methods previously established
  for ${\cal N}=0$ supergravity. As a cross-check, we rederive the
  same formulae using a tensorial approach, which then allows us to
  extend our arguments to sources containing arbitrary powers of the
  inverse radial coordinate. We also generalise our results to include
  the Kerr-Newman black hole, clarifying previous alternative double
  copy formulae presented in the literature. Our results extend the
  validity of the Weyl double copy, and may be useful for further
  astrophysical applications of this correspondence.
\end{abstract}

\vspace*{0.5cm}

\section{Introduction}
\label{sec:intro}

Recent years have seen intense interest in the relationships between
different field theories, such as the gauge and gravity theories
underlying fundamental interactions in our universe. One such
relationship is the {\it double copy}, which first arose in the study
of scattering amplitudes~\cite{Bern:2010ue,Bern:2010yg}, itself
inspired by previous results in string
theory~\cite{Kawai:1985xq}. This states that certain quantities in
gauge theories can be straightforwardly recycled to produce gravity
results, and it is by now well-known that the idea can be applied to
classical solutions of each theory, where certain examples can be
mapped
exactly~\cite{Monteiro:2014cda,Luna:2015paa,Ridgway:2015fdl,Bahjat-Abbas:2017htu,Carrillo-Gonzalez:2017iyj,CarrilloGonzalez:2019gof,Bah:2019sda,Alkac:2021seh,Alkac:2022tvc,Luna:2018dpt,Sabharwal:2019ngs,Alawadhi:2020jrv,Godazgar:2020zbv,White:2020sfn,Chacon:2020fmr,Chacon:2021wbr,Chacon:2021hfe,Chacon:2021lox,Dempsey:2022sls,Easson:2022zoh,Chawla:2022ogv,Han:2022mze,Armstrong-Williams:2022apo,Han:2022ubu}. Even
if this is not the case, perturbative classical double copies are
possible~\cite{Elor:2020nqe,Farnsworth:2021wvs,Anastasiou:2014qba,LopesCardoso:2018xes,Anastasiou:2018rdx,Luna:2020adi,Borsten:2020xbt,Borsten:2020zgj,Goldberger:2017frp,Goldberger:2017vcg,Goldberger:2017ogt,Goldberger:2019xef,Goldberger:2016iau,Prabhu:2020avf,Luna:2016hge,Luna:2017dtq,Cheung:2016prv,Cheung:2021zvb,Cheung:2022vnd,Cheung:2022mix,Chawla:2023bsu,Easson:2023dbk,Farnsworth:2023mff,Borsten:2023paw,Alkac:2023glx,He:2023iew,Chawla:2024mse},
which have direct applications to gravitational wave physics. It is
not currently known how general the double copy is, and in particular
whether it extends beyond (weak coupling) perturbation theory (see
e.g. refs.~\cite{Monteiro:2011pc,Borsten:2021hua,Alawadhi:2019urr,Banerjee:2019saj,Huang:2019cja,Berman:2018hwd,Alfonsi:2020lub,Alawadhi:2021uie,White:2016jzc,DeSmet:2017rve,Bahjat-Abbas:2018vgo,Cheung:2022mix,Moynihan:2021rwh,Borsten:2022vtg,Armstrong-Williams:2024icu}
for non-perturbative aspects). Recent reviews of the double copy may
be found in
refs.~\cite{Borsten:2020bgv,Bern:2019prr,Adamo:2022dcm,Bern:2022wqg,White:2021gvv,White:2024pve}.

Given the above remarks, there is clear scope for studies which aim to
broaden the remit of the double copy, by studying cases or examples
which go beyond what has been studied before. One such frontier
consists of non-vacuum classical solutions of gauge theories or
gravity. A number of exact classical double copies are known, such as
the Kerr-Schild double copy of ref.~\cite{Monteiro:2014cda}, the Weyl
double copy of ref.~\cite{Luna:2018dpt}, and the twistor double copy
of refs.~\cite{White:2020sfn,Chacon:2021wbr} (see
ref.~\cite{Luna:2022dxo} for a discussion of how these various schemes
are related). In each of the original incarnations, only certain
vacuum solutions were considered, and the question then naturally
arises as to whether such conditions can be relaxed. Early work on the
inclusion of sources in the Kerr-Schild double copy can be found in
ref.~\cite{Ridgway:2015fdl}. More recently,
refs.~\cite{Easson:2021asd,Easson:2022zoh} have looked at non-vacuum
solutions in the context of the Weyl double copy of
ref.~\cite{Luna:2018dpt}, which we will review below. The authors of
refs.~\cite{Easson:2021asd,Easson:2022zoh} found various relationships
between particular non-vacuum gauge and gravity solutions. Explicit
examples included those in Einstein-Maxwell theory, in which General
Relativity is coupled to a photon, and whose solutions include the
well-known Kerr-Newman black hole. The double copy formulae presented
in refs.~\cite{Easson:2021asd,Easson:2022zoh} take the form of product
relationships between spinor fields in electromagnetic theories, and
their counterparts in gravity. Given the particular nature of the
solutions considered, however, it is not yet clear whether such
formulae are generally applicable to all solutions of these
theories. This motivates trying to derive such formulae from first
principles, which is the subject of this paper.

A general prescription for deriving Weyl double copy formulae has been
developed in
refs.~\cite{White:2020sfn,Chacon:2021wbr,Chacon:2021lox,Chacon:2021hfe,Luna:2022dxo,CarrilloGonzalez:2022ggn,Armstrong-Williams:2023ssz},
also taking inspiration from ref.~\cite{Guevara:2021yud}. By starting
with solutions of classical equations in momentum space, one may
perform an inverse Fourier transform to position space, but where one
splits the inverse transform into two stages. The first takes one into
an intermediate twistor space, such that classical solutions are
represented by certain functions there. The second part of the
transform then corresponds to the well-known Penrose
transform~\cite{Penrose:1967wn,Penrose:1972ia} that relates twistor
space functions (strictly speaking, representatives of cohomology
classes~\cite{Eastwood:1981jy}) to spinorial fields in spacetime. The
latter can then be seen to obey the Weyl double copy where relevant,
and thus this recipe can be used to ``derive'' Weyl double copy
formulae whenever they exist. Applications have so far included the
original Weyl double copy of ref.~\cite{Luna:2018dpt}, a novel variant
known as the Cotton double copy~\cite{Emond:2022uaf,Gonzalez:2022otg}
in three spacetime dimensions~\cite{CarrilloGonzalez:2022ggn} (see also refs.\cite{Moynihan:2020ejh,Gonzalez:2021bes,Burger:2021wss,Gonzalez:2021ztm}), and an
extension of the double copy that applies to ${\cal N}=0$
supergravity~\cite{Armstrong-Williams:2023ssz}. Here, we will adapt
this scheme to probe Weyl double copies for sourced fields, finding a
number of complications with respect to previous cases. However, we
will succeed in deriving Weyl double copy formulae for a particular
non-zero source. We will then cross-check our results by showing how
they can be obtained starting from the conventional tensorial
formulation of field theory, which will also allow us to generalise
our result to arbitrary spherically symmetric sources. Next, we show
that certain axially symmetric examples from
ref.~\cite{Easson:2022zoh} can also be obtained, such as the
Kerr-Newman black hole \cite{Moynihan:2019bor}. We furthermore clarify the relationship
between apparently different doubly copy formulae presented there. We
hope that our results prove useful in extending the double copy, and
also provide useful data for future studies in this area.

The structure of our paper is as follows. In section~\ref{sec:review},
we review the Weyl double copy, and also the conjectures of
refs.~\cite{Easson:2021asd,Easson:2022zoh} regarding how it may extend
to sourced fields. In section~\ref{sec:source}, we take a particular
type of source, corresponding to the Reissner-Nordstrom black hole. We
show how to extend our previous formalism for deriving Weyl double
copies, yielding to explicit spacetime formulae linking gauge and
gravity results. In section~\ref{sec:tensor}, we derive similar
results using tensor methods, and also argue how they generalise to
more complicated sources. We examine the Kerr-Newman black hole in
section~\ref{sec:spin}, before discussing our results and concluding
in section~\ref{sec:conclude}.

\section{Review of the Weyl double copy}
\label{sec:review}

In this section, we review salient details regarding the Weyl double
copy, which are needed for what follows. Our starting point is the
{\it spinorial formalism} of field theory, which states that all
tensorial quantities such as field strengths and curvature tensors can
be converted into multi-index spinor objects via contraction with {\it
  Infeld-van-der-Waerden symbols}. If we are working in Minkowski
spacetime with
line element
\begin{equation}
  ds^2=dt^2-dx^2-dy^2-dz^2,
  \label{ds213}
\end{equation}
we will follow the conventions of ref.~\cite{Guevara:2021yud} in
choosing
\begin{equation}
  \sigma^\mu_{A\dot{A}}=\left(I,\sigma_z,\sigma_x,\sigma_y\right)_{A\dot{A}},
  \label{sigmadef}
\end{equation}
where $I$ is the identity matrix in spinor space, and $\{\sigma_i\}$
are the Pauli matrices. We will also have occasion to work in (2,2)
signature, with line element
\begin{equation}
  ds^2=dt^2-dx^2-dy^2+dz^2,
  \label{ds222}
\end{equation}
such that eq.~(\ref{sigmadef}) is modifed to
\begin{equation}
  \sigma^\mu_{A\dot{A}}=\left(I,\sigma_z,\sigma_x,i\sigma_y\right)_{A\dot{A}}.
  \label{sigmadef2}
\end{equation}
Whereas tensor indices are raised and lowered with the metric
$\eta^{\mu\nu}$, spinorial indices can be raised / lowered with the
Levi-Civita symbols
\begin{align}
\epsilon_{A B} =
\begin{pmatrix}
0 & 1 \\
-1 & 0 
\end{pmatrix},\quad 
\epsilon^{AB} = 
\begin{pmatrix}
0&1\\
-1&0 
\end{pmatrix},
\label{levi22b}
\end{align}
according to e.g.
\begin{align}
  q_{A} =  q^{B} \epsilon_{B A},\quad
  q^{A} = \epsilon^{AB} q_{B} .
  \label{raiselower}
\end{align}
Upon translating to the spinor language, an $n$-index tensor
$V_{\alpha\beta\ldots\gamma}$ is associated with a $2n$-index spinor:
\begin{equation}
  V_{\alpha\beta\ldots\gamma}\sigma^\alpha_{A\dot{A}}
  \sigma^{\beta}_{B\dot{B}}\ldots
  \sigma^{\gamma}_{C\dot{C}}=V_{A\dot{A}B\dot{B}\ldots C\dot{C}},
  \label{Vtranslate}
\end{equation}
where we keep the symbol $V$ for the tensor or spinor quantity, given
that the nature of indices makes this unambiguous. For spacetime
vectors, we note the useful identity
\begin{equation}
  V\cdot W=\frac12 V^{A\Ad}W_{A\Ad}.
  \label{xdoty}
\end{equation}
Also, the spinorial translation of a spacetime vector satisfies
\begin{equation}
  {\rm det}\left[V_{A\Ad}\right]=V^2,
  \label{detV}
\end{equation}
which vanishes for a null vector, such that its spinorial translation
may be written as an outer product of two spinors:
\begin{equation}
  V_{A\Ad}=\eta_A\tilde{\eta}_{\Ad}.
  \label{outerprod}
\end{equation}
Armed with these definitions, the familiar tensor quantities of field
theory can all be translated into multi-index spinors. This turns out
to be advantageous in many circumstances due to two particularly
useful properties. First, a general multi-index spinor can always be
decomposed into sums of products of spinors which are fully symmetric
in their indices, and Levi-Civita tensors. For example, the field
strength tensor $F_{\mu\nu}$ yields a spinor field
$F_{A\dot{A}B\dot{B}}$, which can be decomposed
as~\cite{Penrose:1987uia,Stewart:1990uf}
\begin{equation}
  F_{A\dot{A}B\dot{B}}(x)=\Phi_{AB}(x)\epsilon_{\Ad\Bd}
  +\bar{\Phi}_{\Ad\Bd}(x)\epsilon_{AB}.
  \label{Fdecomp}
\end{equation}
Physically, $\phi_{AB}$ and $\phi_{\Ad\Bd}$ represent the
self-dual and anti-self-dual parts of the electromagnetic field respectively. In
Lorentzian signature, these quantities are related by complex
conjugation, whereas they constitute independent real degrees of
freedom in split signature. Gravity is described in terms of the
Riemann curvature tensor, whose spinorial translation has the
decomposition~\cite{Witten:1959zza,Penrose:1987uia,Stewart:1990uf}
\begin{equation}
\begin{gathered}
  R_{A B C D \Ad \Bd \Cd \Dd}=\Psi_{A B C D} \epsilon_{\Ad
    \Bd} \epsilon_{\Cd \Dd}+\bar{\Psi}_{\Ad \Bd \Cd \Dd}
  \epsilon_{A B} \epsilon_{C D}+\Phi_{A B \Cd \Dd}
  \epsilon_{\Ad \Bd} \epsilon_{C D}+\bar{\Phi}_{\Ad \Bd C D} \epsilon_{A B} \epsilon_{\Cd \Dd} \\
		+2 \Lambda\left(\epsilon_{A C} \epsilon_{B D} \epsilon_{\Ad \Cd} \epsilon_{\Bd \Dd}-\epsilon_{A D} \epsilon_{B C} \epsilon_{\Ad \Dd} \epsilon_{\Bd \Cd}\right).
\end{gathered}	  
\label{Rdecomp}
\end{equation}
For vacuum solutions, all but the first two terms on the right-hand
side are absent, and the Riemann tensor itself reduces to the {\it
  Weyl tensor} $C_{\alpha\beta\gamma\delta}$, with spinorial
translation
\begin{equation}
\begin{gathered}
  C_{A B C D \Ad \Bd \Cd \Dd}=\Psi_{A B C D} \epsilon_{\Ad
    \Bd} \epsilon_{\Cd \Dd}+\bar{\Psi}_{\Ad \Bd \Cd \Dd}
  \epsilon_{A B} \epsilon_{C D}.
\end{gathered}	  
\label{Cdecomp}
\end{equation}
Analogous to the case of electromagnetism, the quantity $\Psi_{ABCD}$
($\bar{\Psi}_{\Ad\Bd\Cd\Dd}$) describes the (anti-)self-dual part of
the field respectively, and is known as the (conjugate) Weyl spinor.

The second useful property in the spinorial formalism is that fully
symmetric spinors can themselves be decomposed in terms of
single-index {\it principal spinors}. For example, a given Weyl spinor
may be decomposed as
\begin{equation}
  \Psi_{ABCD}=\alpha_{(A}\beta_{B}\gamma_C\delta_{D)}.
\label{Weyldecomp}
\end{equation}
Some or all of the principal spinors may be mutually proportional, and
the different patterns of degeneracy are known as {\it Petrov
  types}. The Weyl double copy can now be phrased as follows. For
certain vacuum solutions of electromagnetism and gravity, the Weyl
spinor of the latter can be expressed as
\begin{equation}
  \Psi_{ABCD}=\frac{\Phi_{(AB}\Phi_{CD)}}{\phi},
  \label{WeylDC}
\end{equation}
where the brackets denote symmetrisation over indices, $\phi_{AB}$ is
the relevant electromagnetic spinor, and $\phi$ is a solution of a
linearised scalar theory satisfying
\begin{equation}
  \partial^2\phi=0.
  \label{scallin}
\end{equation}
The corresponding relationship for the conjugate Weyl spinor can be
obtained by replacing all quantities in eq.~(\ref{WeylDC}) with their
conjugates. Equation~(\ref{WeylDC}) was argued to hold for general
vacuum Petrov type D solutions in ref.~\cite{Luna:2018dpt}, and
extended to type N solutions in ref.~\cite{Godazgar:2020zbv}. Already
in ref.~\cite{Luna:2018dpt}, the possibility of mixed Weyl double copy
formulae was given, namely that one may take a pair of different
electromagnetic spinors and combine them according to
\begin{equation}
  \Psi_{ABCD}=\frac{\Phi^{(1)}_{(AB}\Phi^{(2)}_{CD)}}{\phi},
  \label{WeylDC2}
\end{equation}
also yielding a valid Weyl spinor. In
refs.~\cite{White:2020sfn,Chacon:2021wbr}, a derivation of the Weyl
double copy was given using ideas from twistor theory, which also
showed that the formalism could be generalised to other Petrov types,
albeit at linearised level in the vacuum field
equations. Reference~\cite{Luna:2022dxo}, inspired by
ref.~\cite{Guevara:2021yud}, clarified matters further by showing that
the momentum-space amplitudes double copy amounts to the same thing as
the twistor and Weyl approaches. Furthermore, these three double
copies can be systematically related to each other by well-defined
integral transforms. We will see these arguments in detail in
section~\ref{sec:source}, and indeed generalise them. For now, we note
that a similar scheme has been used to derive analogues of the Weyl
double copy for vacuum solutions of topologically massive gauge and
gravity theories in three
dimensions~\cite{CarrilloGonzalez:2022ggn,Emond:2022uaf,Gonzalez:2022otg},
and for ${\cal N}=0$ supergravity. In the latter, the generalised
Riemann tensor has a decomposition similar to eq.~(\ref{Rdecomp}),
which contains mixed-index spinors in addition to the Weyl spinors. We
will refer to $\Phi_{AB\Cd\Dd}$ as the {\it Ricci spinor} in what
follows.

The presence of sources in the Weyl double copy was first considered
in refs.~\cite{Easson:2021asd,Easson:2022zoh}, where the basic idea is
to replace eq.~(\ref{WeylDC}) with a series of terms:
\begin{equation}
  \Psi_{ABCD}=\sum_{n=1}^m\frac{1}{\phi^{(n)}}\Phi^{(n)}_{(AB}
  \Phi^{(n)}_{CD)},
  \label{MantonDC}
\end{equation}
and similarly for the conjugate Weyl spinor. The right-hand side
contains a tower of electromagnetic spinors $\Phi^{(n)}_{AB}$ and
scalar fields $\phi^{(n)}$, such that the $n=1$ term corresponds to
the traditional vacuum Weyl double copy. Each $n>1$ electromagnetic
spinor corresponds to a field strength $F_{\mu\nu}^{(n)}$ satisfying a
non-vacuum Maxwell equation
\begin{equation}
  \partial^\mu F_{\mu\nu}^{(n)}=j^{(n)}_\mu.
  \label{Fmununonvac}
\end{equation}
Likewise, each scalar field $\phi^{(n)}$ satisfies a non-vacuum
equation
\begin{equation}
  \partial^2 \phi^{(n)}=\rho_S^{(n)}
  \label{Snonvac}
\end{equation}
for some charge density $\rho_S^{(n)}$. The physical interpretation of
eq.~(\ref{MantonDC}) is then that one has decomposed a given gravity
solution into $n$ distinct terms, each of which has a well-defined
single and zeroth copy. One particular example in
refs.~\cite{Easson:2021asd,Easson:2022zoh} consists of expanding a
given gravity solution in inverse powers of the radial coordinate
$r$. Each power then corresponds to a distinct term in
eq.~(\ref{MantonDC}). Another example concerns Einstein-Maxwell
theory, namely General Relativity coupled to electromagnetism. For
certain solutions, one may truncate the sum in eq.~(\ref{MantonDC}) at
$n=2$:
\begin{equation}
  \Psi_{ABCD}=\frac{1}{\phi^{(1)}}\Phi^{(1)}_{(AB}\Phi^{(1)}_{CD)}
    +\frac{1}{\phi^{(2)}}\Phi^{(2)}_{(AB}\Phi^{(2)}_{CD)}.
      \label{EinsteinMax}
\end{equation}
Consistent with the above remarks, the gauge field $A_\mu^{(1)}$
corresponding to $\phi^{(1)}_{AB}$ satisfies a vacuum equation, as
does the scalar field $\phi^{(1)}$. However, the fields $A_\mu^{(2)}$
and $\phi^{(2)}$ will satisfy non-vacuum equations, where the
corresponding charge density and currents must be physically analogous
to the source in the gravity theory. An explicit case is that of the
Kerr-Newman black hole, for which one finds (up to an overall constant
factor in each case)~\cite{Easson:2022zoh}
\begin{align}
  \phi^{(1)}=\frac{1}{r+ia\cos\theta},\quad
  \Phi^{(1)}_{AB}=\frac{1}{(r+ia\cos\theta)^2} o_{(A}\imath_{B)}
\label{KN1}
\end{align}
and
\begin{align}
  \phi^{(2)}=\frac{1}{(r+ia\cos\theta)(r-ia\cos\theta)},\quad
  \Phi^{(2)}_{AB}=\frac{1}{(r+ia\cos\theta)^2(r-ia\cos\theta)}
  o_{(A}\imath_{B)}.
\label{KN2}
\end{align}
Here $r$ and $\theta$ are radial and polar angular coordinates, and
$a$ a constant parameter related to the angular momentum of the black hole in gravity.

The corresponding scalar, electromagnetic and gravitational charge /
energy densities are then found to be
\begin{equation}
  \rho^{(2)}_S\propto \rho_e^{(2)}\propto \rho_{\rm grav.}^{(2)}
  \propto \frac{(r^2+a^2)+a^2\sin^2\theta}{\rho^6},\quad
  \rho^2=r^2+a^2\cos^2\theta.
  \label{rhoKN}
\end{equation}
For a non-vacuum solution, the Riemann tensor will have additional
components in addition to the Weyl tensor. In terms of the spinorial
decomposition of eq.~(\ref{Rdecomp}), this means that the mixed-index
spinors will be nonzero. Reference~\cite{Easson:2022zoh} then
conjectured the following double copy formula for the Kerr-Newman
solution:
\begin{equation}
  \Phi_{AB\Cd\Dd}=\frac{1}{\phi^{(2)}}\Phi^{(2)}_{AB}
  \bar{\Phi}^{(2)}_{\Cd\Dd}.
  \label{Mantonmixed}
\end{equation}
An alternative formula was also given in ref.~\cite{Easson:2022zoh},
that expresses the Ricci spinor as a product of the vacuum-like
electromagnetic spinors:
\begin{equation}
  \Phi_{AB\Cd\Dd}\propto\Phi^{(1)}_{AB}
  \bar{\Phi}^{(1)}_{\Cd\Dd}.
  \label{Mantonmixed2}
\end{equation}
This form was argued to be more generally applicable, including to
spacetimes that are non-asymptotically flat. Given the previous
studies that have derived (generalised) Weyl double copies from first
principles~\cite{Luna:2022dxo,CarrilloGonzalez:2022ggn,Armstrong-Williams:2023ssz},
we may now ask whether similar methods may be used in deriving sourced
Weyl double copy formulae directly. We commence this investigation in
the following section.

\section{A Weyl double copy for the Reissner-Nordstrom black hole}
\label{sec:source}

According to eq.~(\ref{MantonDC}), a sourced Weyl double copy states
that a gravitational Weyl spinor can be written as a sum of products
of electromagnetic spinors, where each term has a certain scalar field
divided out. In order to probe this relationship, we may take a single
well-defined term in such a sum, and then ascertain whether or not we
can reproduce it using similar methods to those used for other
Weyl(-like) double
copies~\cite{Luna:2022dxo,CarrilloGonzalez:2022ggn}. That is, we may
pick a set of scalar, electromagnetic and graviton fields whose
sources are physically related as follows:
\begin{equation}
  \rho_S=\rho(x),\quad j^\mu(x)=\rho(x)u^\mu,\quad T^{\mu\nu}(x)=
  \rho(x)u^\mu u^\nu,
  \label{rhoSdef}
\end{equation}
where we will assume static and spherically symmetric sources, such
that $u^\mu=\delta^\mu_0$ is a vector pointing in the time
direction. In the scalar and gauge theories, $\rho(x)$ represents the
appropriate charge density, whereas it will be an energy density in
gravity. We have furthermore ignored dimensionful coupling constants
in our definitions, for brevity. Reference~\cite{Luna:2022dxo}
provided a systematic derivation of the Weyl double copy using the
following three-step procedure:
\begin{enumerate}
  \item[(i)] One may express spinorial solutions of (linearised)
    classical equations for scalar, gauge and gravity theory as
    inverse Fourier transforms of momentum-space solutions.
  \item[(ii)] One may transform the momentum integral to certain
    spinor variables, and carry out some of the integrals to yield an
    intermediate twistor-space representation of each classical
    solution. This step reproduces the twistor double copy first
    explored in refs.~\cite{White:2020sfn,Chacon:2021wbr}.
  \item[(iii)] The remaining integral can be carried out to yield
    position-space representations of each classical solution, with
    their principal spinors clearly identified. These solutions are
    then found to obey the Weyl double copy.
\end{enumerate}
Here, we will apply a similar scheme to sourced solutions, for which
we must begin by expressing classical solutions in terms of momentum
space integrals. Due to the static nature of each source in
eq.~(\ref{rhoSdef}), their Fourier transforms may be written as
\begin{equation}
  \tilde{\rho}_S(k)=\delta(u\cdot k){\cal J}(\vec{k}),\quad
  \tilde{j}^\mu(k)=\delta(u\cdot k){\cal J}(\vec{k})u^\mu,\quad
  \tilde{T}^{\mu\nu}(k)=
  \delta(u\cdot k){\cal J}(\vec{k})u^\mu u^\nu,
  \label{rhoSdef2}
\end{equation}
where we have introduced the 4-momentum $k^\mu=(k^0,\vec{k})$, and
noted that $k^0=u\cdot k$. The classical solution for the scalar field
given the source in eq.~(\ref{rhoSdef}) can then be written as
\begin{equation}
  \phi(x)=\int\frac{d^4 k}{(2\pi)^4}\delta(u\cdot k)
  \frac{{\cal J}(\vec{k})}{k^2}e^{-ik\cdot x}.
  \label{phisol}
\end{equation}
Likewise, the gauge field (in Lorenz gauge $\partial\cdot A=0$) has
the form
\begin{equation}
  A^\mu(x)=\int\frac{d^4 k}{(2\pi)^4} u^\mu \delta(u\cdot k)
  \frac{{\cal J}(\vec{k})}{k^2}e^{-ik\cdot x},
  \label{Amusol}
\end{equation}
leading to the gauge-invariant field-strength tensor
\begin{equation}
  F^{\mu\nu}(x)=\int\frac{d^4 k}{(2\pi)^4} k^{[\mu}u^{\nu]} \delta(u\cdot k)
  \frac{{\cal J}(\vec{k})}{k^2}e^{-ik\cdot x}.
  \label{Amusol2}
\end{equation}
Finally, we must consider the corresponding gravity solution, which we
may assume to be a solution of the linearised Einstein equations. The
linearised Riemann tensor can be expressed in terms of the graviton as
\[
R^{\mu \nu \rho \sigma}(x)=\frac{\kappa}{2}\left(\partial^\sigma \partial^{[\mu} h^{\nu] \rho}+\partial^\rho \partial^{[\nu} h^{\mu] \sigma}\right),
\]
where the graviton itself is given by
\[
h^{\mu\nu}(x) - \frac12\eta^{\mu\nu}h(x) = \int \frac{d^4 k}{(2\pi)^4}
\frac{\tilde{T}^{\mu\nu}(k)}{k^2}e^{-ik\cdot x}.
\]
Taking the trace and inverting, one finds
\[
h^{\mu\nu}(x) = \int \frac{d^4 k}{(2\pi)^4}
\frac{\tilde{T}^{\mu\nu}(k) - \frac12\eta^{\mu\nu}
  \tilde{T}(k)}{k^2}e^{-ik\cdot x},
\]
and using this result together with the source of eq.~(\ref{rhoSdef2})
yields
\begin{equation}
  R^{\mu\nu\rho\sigma}(x)=\kappa\int\frac{d^4k}{(2\pi)^4}
  \left[U^{\rho[\mu}k^{\nu]}k^\sigma-U^{\sigma[\mu}k^{\nu]}k^\rho\right]
  \frac{{\cal J}(\vec{k})}{k^2}e^{-ik\cdot x},
  \label{Rsol}
\end{equation}
where
\[
U^{\nu\rho} = u^\nu u^\rho - \frac12\eta^{\nu\rho}.
\]
In the above results, we can recognise that the electromagnetic and
gravitational expressions can be rewritten in terms of derivative
operators acting on the scalar field $\phi$. In particular, comparison
of eqs.~(\ref{Amusol2}) and eq.~(\ref{phisol}) shows that the field
strength tensor associated with a given scalar solution is given by
\[
  \label{Fphi}
F_{\mu\nu}^{(n)} = 2u_{[\mu}\pd_{\nu]}\phi^{(n)}.
\]
Furthermore, the graviton and Riemann curvature tensor can be written
as
\[
h^{(n)}_{\mu\nu} = \left(u_\mu u_\nu-\frac12\eta_{\mu\nu}\right)\phi^{(n)}(r).
\label{hphi}
\]
and
\[
R^{(n)}_{\mu\nu\rho\sigma}(x) &= -2\left[U_{\rho[\mu}\pd_{\nu]}\pd_\sigma - U_{\sigma[\mu}\pd_{\nu]}\pd_\rho\right]\phi^{(n)}(r).
\label{Rphi}
\]
Having seen how to obtain the tensor forms of various quantities, we next
need to translate them into the spinor formalism, which is the subject
of the following section.

\subsection{Scalar field}
\label{sec:scalarRN}

Considering the scalar momentum integral of eq.~(\ref{phisol}), we can
introduce a lightcone decomposition 
\begin{equation}
k^\mu = \omega\ell^\mu + \xi q^\mu,
\label{kdecomp}
\end{equation}
where $\ell^2 = q^2 = 0$; the parameters $\omega$ and $\xi$ have
dimensions of energy; and
\begin{equation}
  q^\mu=\frac12(0,0,-1,1)
  \label{qdef}
\end{equation}
is a fixed 4-vector. From eq.~(\ref{outerprod}), the spinorial
translations of $l^\mu$ and $q^\mu$ will each decompose into an outer
product of two spinors, so that one may write
\begin{equation}
  k_{A\Ad}=\omega \lambda_A\tilde{\lambda}_{\Ad}+\xi q_A\tilde{q}_{\Ad}.
  \label{kAAd}
\end{equation}
The delta function in eq.~(\ref{phisol}) imposes the condition
\begin{equation}
  u\cdot k=0\quad\Rightarrow\quad u\cdot l=0,
  \label{udotk2}
\end{equation}
where the second condition follows from eq.~(\ref{kdecomp}), and the
fact that our choice of $q^\mu$ implies $u\cdot q=0$. In the spinor
language, this amounts to
\begin{equation}
  u^{A\Ad}\lambda_A\tilde{\lambda}_{\Ad}=0,
  \label{udotkspin}
\end{equation}
which in turn implies
\begin{equation}
  \lambda_A\propto {u_A}^{\Ad}\tilde{\lambda}_{\Ad}.
  \label{lambdaprop}
\end{equation}
We can turn this into an equality by parametrising the spinors
appearing in eq.~(\ref{kdecomp}) as follows:
\begin{equation}
  \lambda_A=\left(\begin{array}{c}\sqrt{1/z}\\\sqrt{z}\end{array}\right),\quad
  \tilde{\lambda}_{\Ad}=\left(\begin{array}{c}\sqrt{-1/\tilde{z}}
    \\-\sqrt{-\tilde{z}}\end{array}\right).
  \label{lamparam}
\end{equation}
To see this, note that an explicit calculation gives
\begin{equation}
  \delta(u\cdot k)=\frac{2}{\omega}\delta\left(\frac{1+z\tilde{z}}
        {\sqrt{-z\tilde{z}}}\right),
        \label{deltafn1}
\end{equation}
and hence
\begin{equation}
  z=-\frac{1}{\tilde{z}}.
  \label{zztilde}
\end{equation}
From eq.~(\ref{lamparam}), one may also calculate
\begin{equation}
  {u_A}^{\Ad}\tilde{\lambda}_{\Ad}=\left(\begin{array}{c}
    \sqrt{-\tilde{z}}\\ 1/\sqrt{-\tilde{z}}\end{array}\right),
  \label{ulam}
\end{equation}
which indeed yields $\lambda_A$ upon using
eq.~(\ref{zztilde}). Returning to eq.~(\ref{phisol}), we can now
transform the momentum integration to the variables
$(\omega,\xi,z,\tilde{z})$. The Jacobian is found to be
\begin{equation}
  J=\frac{\omega^2\sqrt{-\tilde{z}}}{4\tilde{z}^2\sqrt{z}},
  \label{Jdef}
\end{equation}
and one also finds 
\begin{equation}
  k^2=-|\vec{k}|^2=\frac{\xi\omega}{\tilde{z}},
  \label{k2res}
\end{equation}
such that eq.~(\ref{phisol}) becomes
\begin{equation}
  \phi(x)=\int\frac{d\omega d\xi dz d\tilde{z}}
      {2(2\pi)^4}\frac{\omega\sqrt{-\tilde{z}}}{\tilde{z}\sqrt{z}}
      \delta\left(\frac{1+z\tilde{z}}{\sqrt{-z\tilde{z}}}\right)
      e^{-i\xi q\cdot x}\exp\left[-\frac{i\omega}{2}\tilde{\lambda}_{\Ad}
        \tilde{\lambda}_{\Bd}x^{A\Ad}{u_A}^{\Bd}\right]
      {\cal J}(\vec{k}).
  \label{phiint}
\end{equation}
A similar integral occured in deriving the original Weyl double copy
in ref.~\cite{Luna:2022dxo}, whose starting point was to note that
certain vacuum classical solutions could be expressed as inverse
Fourier transforms of on-shell momentum-space scattering
amplitudes. In that case, the integral over $\xi$ becomes trivial, in
that $\xi$ is fixed to be zero, corresponding to the on-shellness of
the emitted radiation. This can also be seen from eq.~(\ref{kdecomp}):
if $\xi=0$, then the second term vanishes, and the remaining term is
null. For general non-zero sources, as represented by ${\cal J}(\vec{k})$,
the $\xi$ integral instead becomes non-trivial, corresponding
physically to the fact that one must explicitly account for off-shell
modes of the emitted radiation. This is not a problem in practice, and
to make progress we will study a particular choice for
${\cal J}(\vec{k})$, corresponding to arguably the simplest solution of
Einstein-Maxwell theory, namely the Reissner-Nordstrom black hole. The
scalar field associated with the non-vacuum term is given in the
$a\rightarrow 0$ limit of eq.~(\ref{KN2}), as
\begin{equation}
  \phi^{(2)}\propto\frac{1}{r^2},
  \label{phiKN}
\end{equation}
One then finds
\begin{equation}
  \partial^2\phi^{(2)}\propto\frac{1}{r^4}\quad\Rightarrow\quad \rho_S(x)\propto
  \frac{1}{r^4}.
  \label{rhoSKN}
\end{equation}
In momentum space, this implies that we consider a source term
\begin{equation}
  {\cal J}(\vec{k})= 2\pi^2|\vec{k}|.
  \label{JRN}
\end{equation}
Substituting this into eq.~(\ref{phiint}) and using eq.~(\ref{k2res}),
one may straightforwardly evaluate the $\xi$, $\omega$ and $z$
integrals to obtain
\begin{equation}
  \phi^{(2)}(x)=\frac{1}{\sqrt{32}\pi}\frac{1}{\sqrt{q\cdot x}}\int d\tilde{z}
  \frac{1}{\tilde{z}^{3/2}}\frac{1}{\left(\tilde{\lambda}_{\Ad}
    \tilde{\lambda}_{\Bd}x^{A\Ad}{u_A}^{\Bd}\right)^{3/2}}.
  \label{phiint2}
\end{equation}
It is now convenient to define the rescaled spinor
\begin{equation}
  \tilde{\chi}_{\Ad}=\sqrt{-\tilde{z}}\tilde{\lambda}_{\Ad}=
  \left(\begin{array}{c}
    1\\\tilde{z}
  \end{array}\right),
  \label{chidef}
\end{equation}
such that eq.~(\ref{phiint2}) can be written more compactly as
\begin{equation}
  \phi^{(2)}(x)=\frac{1}{\sqrt{32}\pi}\frac{1}{\sqrt{q\cdot x}}\int d\tilde{z}
  \frac{1}{\left(\tilde{\chi}_{\Ad}
    \tilde{\chi}_{\Bd}x^{A\Ad}{u_A}^{\Bd}\right)^{3/2}}.
  \label{phiint3}
\end{equation}
Similar integrals were obtained in the derivation of the Weyl double
copy of ref.~\cite{Luna:2022dxo}, where they could be interpreted in
projective twistor space $\mathbb{PT}$. Points in $\mathbb{PT}$
consist of a pair of spinors collected into a single 4-component
twistor
\begin{equation}
  Z^\alpha=\left(\omega^A,\tilde{\chi}_{\Ad}
  \right),
  \label{twistor}
\end{equation}
where the two spinors satisfy the so-called {\it incidence relation}
\begin{equation}
  \omega^A=x^{A\Ad}\tilde{\chi}_{\Ad}.
  \label{incidence}
\end{equation}
This relation is invariant under a simultaneous (complex) rescaling of
both sides, which amounts to an overall scaling $Z^\alpha\rightarrow
\alpha Z^\alpha$ in eq.~(\ref{twistor}). It is for this reason that
twistors obeying the incidence relation are associated with projective
twistor space. With the parametrisation of eq.~(\ref{chidef}), the
measure in eq.~(\ref{phiint3}) can be written in the invariant form
\begin{equation}
  d\tilde{z}=\tilde{\chi}_{\Bd}d\tilde{\chi}^{\Bd}.
  \label{measure}
\end{equation}
Then eq.~(\ref{phiint3}) can be written in terms of twistor variables as
\begin{equation}
  \phi^{(2)}(x)=\frac{\lambda}{\sqrt{q\cdot x}}\int \tilde{\chi}_{\Bd}d\tilde{\chi}^{\Bd}
  \frac{1}{\left(Q_{\alpha\beta} Z^\alpha Z^\beta\right)^{3/2}},\quad
  Q_{\alpha\beta}=\left(\begin{array}{cc}0 & {u_A}^{\Bd}\\ {u_A}^{\Bd} & 0
    \end{array}\right).
  \label{phiint4}
\end{equation}
However, unlike the previous cases of
refs.~\cite{Luna:2022dxo,Armstrong-Williams:2023ssz}, the integral of
eq.~(\ref{phiint4}) cannot be consistently interpreted as a contour
integral in projective twistor space, as the combined measure and
integrand is not invariant under rescalings of $Z^\alpha$. This is not
unexpected, given that the conventional Penrose transform is for
vacuum solutions only. Furthermore, we will not need to formally
pursue a twistorial interpretation of eq.~(\ref{phiint3}) in what
follows.

Another complication of eq.~(\ref{phiint3}) compared to the related
integrals in refs.~\cite{Luna:2022dxo,Armstrong-Williams:2023ssz} is
that its singularity structure involves branch cuts rather than simple
poles, and it is not then immediately clear how to choose an
appropriate contour. However, we can instead carry out the inverse
Fourier transform in eq.~(\ref{phisol}) exactly for our chosen source
to obtain
\begin{align}
  \phi^{(2)}(x)&=\int\frac{d^4 k}{(2\pi)^4}\frac{2\pi^2\delta(u\cdot k)|\vec{k}|}
      {k^2}e^{-ik\cdot x}\\
      &=\frac{1}{(x^2-(u\cdot x)^2)}=\frac{1}{r^2}.
      \label{phiintexact}
\end{align}
Comparison with eq.~(\ref{phiint3}) then yields
\begin{equation}
  \int d\tilde{z}
  \frac{1}{\left(\tilde{\chi}_{\Ad}
    \tilde{\chi}_{\Bd}x^{A\Ad}{u_A}^{\Bd}\right)^{3/2}}=
  \frac{\sqrt{32}\pi\sqrt{q\cdot x}}{r^2}.
    \label{ztildeint}
\end{equation}
It will turn out that all spinor integrals needed for the
electromagnetic and gravitational spinors in subsequent sections can
be expressed in terms of eq.~(\ref{ztildeint}). In anticipation of
this, let us note that the spinor combination appearing in the
denominator of eq.~(\ref{ztildeint}) is given by
\begin{equation}
\tilde{\chi}_{\Ad}
\tilde{\chi}_{\Bd}x^{A\Ad}{u_A}^{\Bd}=\tilde{z}^2(y+z)+2x\tilde{z}
+z-y.
\label{spincomb}
\end{equation}
The integrals we will need to evaluate can then all be written in
terms of the family
\begin{equation}
  I_{m,n}=(-1)^{m+1}\int d\tilde{z}\frac{\tilde{z}^n}
  {[\tilde{z}^2(y+z)+2x\tilde{z}+z-y]^{3/2+m}},
  \label{Imndef}
\end{equation}
where e.g. the scalar field above is given by
\begin{equation}
  \phi^{(2)}(x)=\frac{1}{\sqrt{32}\pi}\frac{1}{\sqrt{q\cdot x}}I_{0,0}.
  \label{phi00}
\end{equation}
with
\begin{equation}
  I_{0,0}=\frac{\sqrt{32}\pi\sqrt{q\cdot x}}{r^2}=\frac{4\pi\sqrt{y+z}}
  {x^2+y^2-z^2}.
  \label{I00}
\end{equation}
By differentiating eq.~(\ref{Imndef}) with respect to the spatial
coordinates $(x,y,z)$, one can derive the recurrence relations
\begin{align}
  \frac{\partial}{\partial x}I_{m,n}=-(3+2m)I_{m+1,n+1},\notag\\
  \left(\frac{\partial}{\partial y}+\frac{\partial}{\partial z}\right)
  I_{m,n}=-(3+2m)I_{m+1,n+2},\notag\\
  \left(\frac{\partial}{\partial y}-\frac{\partial}{\partial z}\right)
  I_{m,n}=(3+2m)I_{m+1,n},
  \label{recurrence}
\end{align}
which we will use repeatedly in what follows.

\subsection{Electromagnetic field}
\label{sec:EM}

Given the scalar field of the previous section, we can consider an
electromagnetic field strength tensor given by eq.~(\ref{Amusol2}),
where we again take the source appearing on the right-hand side to be
given by eq.~(\ref{JRN}). The corresponding electromagnetic spinor can
be obtained by projecting out the relevant components, following
contraction of the tensorial field strength with the
Infeld-van-der-Waerden symbols:
\begin{equation}
  \Phi^{(2)}_{AB}(x)=\epsilon^{\Bd\Ad}\sigma^\mu_{A\Ad}\sigma^\nu_{B\Bd}
  F_{\mu\nu}(x).
  \label{phiABcontract}
\end{equation}
Carrying out the contractions and performing the change of variables
outlined in the previous section yields the following result:
\begin{equation}
  \Phi^{(2)}_{\Ad\Bd}=-\frac{3}{4\pi}
  \frac{1}{(x^{0\dot{1}})^{1/2}}
  \int d\tilde{z}\frac{\tilde{\chi}_{\Ad}
    \tilde{\chi}_{\Bd}}{(\tilde{\chi}_{\Ad}\tilde{\chi}_{\Bd}x^{A\Ad}
    {u_A}^{\Bd})^{5/2}}
  +\frac{1}{4\pi}\frac{1}{(x^{0\dot{1}})^{3/2}}\int d\tilde{z}
  \frac{{u_{\Ad}}^{A}q_A\tilde{q}_{\Bd}+{u_{\Bd}}^{B}q_B\tilde{q}_{\Ad}}
       {(\tilde{\chi}_{\Ad}\tilde{\chi}_{\Bd}x^{A\Ad}
         {u_A}^{\Bd})^{3/2}}.
       \label{phiABintegrand}
\end{equation}
In contrast to previously studied cases, this has two distinct terms,
involving different powers of the spinor combination appearing in the
denominator. Indeed, the second term can be directly traced to the
fact that off-shell modes of the gauge field are included, given that
the vanishing of the $q^\mu$ term in eq.~(\ref{kdecomp}) is associated
with an on-shell gauge field, as remarked above. In order to evaluate
the integrals in eq.~(\ref{phiABintegrand}), one may substitute in the
explicit forms of each spinor, after which one finds that the matrix
form of $\Phi^{(2)}_{AB}$ is 
\begin{equation}
  \Phi^{(2)}_{\Ad\Bd}=\frac{1}{4\pi}\left[\frac{1}{(q\cdot x)^{3/2}}
    \left(\begin{array}{cc}0 & 0\\ 0 & -I_{0,0}\end{array}\right)
    -\frac{6}{\sqrt{q\cdot x}}\left(\begin{array}{cc}I_{1,0} & I_{1,1}\\
      I_{1,1} & I_{1,2}\end{array}\right)
    \right],
  \label{phiABcalc1}
\end{equation}
in terms of the family of integrals introduced in
eq.~(\ref{Imndef}). The recurrence relations of eq.~(\ref{recurrence})
yield
\begin{align}
  I_{1,1}&=-\frac13\frac{\partial}{\partial x}I_{0,0},\notag\\
  I_{1,0}&=\frac13\left(\frac{\partial}{\partial y}-\frac{\partial}
  {\partial z}\right)I_{0,0},\notag\\
  I_{1,2}&=-\frac13\left(\frac{\partial}{\partial y}+\frac{\partial}
  {\partial z}\right)I_{0,0},
  \label{I11}
\end{align}
such that substituting eq.~(\ref{I00}) gives
\begin{align}
  I_{1,0}&=-\frac{32\pi(q\cdot x)^{3/2}}{3r^4},\notag\\
  I_{1,1}&=\frac{16\pi x(q\cdot x)}{3r^4},\notag\\
  I_{1,2}&=\frac{16\pi}{3}(y-z)\frac{\sqrt{q\cdot x}}{r^4}-\frac16 \frac{I_{00}}
  {q\cdot x}.
\label{I11res}
\end{align}
Putting things together, the electromagnetic spinor becomes
\begin{equation}
  \Phi^{(2)}_{AB}=-\frac{2}{r^4}\left(\begin{array}{cc}
    -y-z & x\\ x & -y+z\end{array}\right),
  \label{phiABres}
\end{equation}
which may also be written in the convenient form
\begin{equation}
  \Phi^{(2)}_{AB}=2\frac{{u_{(\Ad}}^B r_{\Bd)B}}{r^4},
  \label{phiABres2}
\end{equation}
where
\begin{equation}
  r_{B\Bd}=(x-(u\cdot x)u)\cdot\sigma_{B\Bd}=\left(\begin{array}{cc}
    -x & z-y \\ -z-y & x\end{array}\right).
    \label{rBB}
\end{equation}

\subsection{Gravitational field}
\label{sec:gravRN}

In the previous two sections, we have assumed a scalar and
electromagnetic source that mimics the source for the
Reissner-Nordstrom black hole. We can now finally calculate the
various gravitational spinors for this solution, and hence ascertain
whether or not there is a non-vacuum generalisation of the Weyl double
copy for this particular solution. First, we recall that the Riemann
curvature for a given source is given by eq.~(\ref{Rsol}). Upon
translating to the spinorial language via
\begin{equation}
  R_{ABCD\Ad\Bd\Cd\Dd}=\sigma^\mu_{A\Ad}\sigma^\nu_{B\Bd}
  \sigma^\rho_{C\Cd}\sigma^\lambda_{D\Dd}R_{\mu\nu\rho\lambda},
  \label{Rspinor}
\end{equation}
one may find the multi-index spinors appearing in the decomposition of
eq.~(\ref{Rdecomp}) via the contractions
\begin{align}
	\Psi_{A B C D} &= \frac{1}{4} R_{(A \dot{X} B}{ }^{\dot{X}}{ }_{C \dot{Y} D)}{ }^{\dot{Y}}\notag\\
	\Phi_{A B \dot{C} \dot{D}} &= \frac{1}{4} R_{(A \dot{X} B)}{ }^{\dot{X}}{ }_{Y (\dot{C} \dot{D})}.
        \label{Rcontract}
\end{align}
For the scalar field $\phi^{(2)}$, carrying out all necessary
contractions and changing to spinor variables as before yields the
result
\[
  \Psi^{(2)}_{\Ad\Bd\Cd\Dd}&=\frac{3\kappa}{2}\int d\tilde{z}\left[
    \frac{5\tilde{\chi}_{\Ad}\tilde{\chi}_{\Bd}\tilde{\chi}_{\Cd}
      \tilde{\chi}_{\Dd}}{(\tilde{\chi}_{\Ad}\tilde{\chi}_{\Bd}x^{A\Ad}
         {u_A}^{\Bd})^{7/2}(q\cdot x)^{1/2}}
    +\frac{2\tilde{\chi}_{(\Ad}\tilde{\chi}_{\Bd}(u\cdot q)_{\Cd} \tilde{q}_{\Dd)}}
    {(\tilde{\chi}_{\Ad}\tilde{\chi}_{\Bd}x^{A\Ad}
      {u_A}^{\Bd})^{5/2}(q\cdot x)^{3/2}}\right.\\
    &\left.\qquad\qquad\qquad+
    \frac{(u\cdot q)_{(\Ad} \tilde{q}_{\Bd}(u\cdot q)_{\Cd} \tilde{q}_{\Dd)}}
    {(\tilde{\chi}_{\Ad}\tilde{\chi}_{\Bd}x^{A\Ad}
      {u_A}^{\Bd})^{3/2}(q\cdot x)^{5/2}}
    \right].
  \label{Psires}
\]
Comparison with the integral form of the electromagnetic spinor in
eq.~(\ref{phiABintegrand}) shows that it is not at all obvious whether
the Weyl spinor of eq.~(\ref{Psires}) is related by any form of double
copy. However, the latter (if it applies) must be true for the
integrated Weyl spinor, and we may carry out the integrals by
expressing them in terms of the master integrals of
eq.~(\ref{Imndef}), and using the recurrence relations of
eq.~(\ref{recurrence}). Using these, one finds that the Weyl spinor requires integrals of the form $I_{2,n}$, which we can related to the lower-spin integrals via
\[
I_{2,n} &= \frac{1}{5}\left(\frac{\pd}{\pd Y}-\frac{\pd}{\pd Z}\right)I_{1,n}\\
&= \frac{1}{15}\left(\frac{\pd}{\pd Y}-\frac{\pd}{\pd Z}\right)^2I_{0,n}.
\]
We find the following required integrals for the Weyl spinor:
\[
I_{2,0} &= \frac{256\pi(q\cdot x)^{5/2}}{15r^6}\\
I_{2,1} &= -\frac{128\pi x(q\cdot x)^{3/2}}{15r^6}\\
I_{2,2} &= \frac{16\pi\sqrt{q\cdot x}}{5r^4} - \frac{128\pi(y-z)(q\cdot x)^{3/2}}{15r^6}\\
I_{2,3} &= \frac{64\pi x(y-z)\sqrt{q\cdot x}}{15r^6} - \frac{8\pi x}{15r^4\sqrt{q\cdot x}}\\
I_{2,4} &= \frac{64\pi (y-z)^2\sqrt{q\cdot x}}{15r^6} - \frac{16\pi (y-z)}{15r^4\sqrt{q\cdot x}} - \frac{2\pi}{15r^2(q\cdot x)^{3/2}}.
\]
We can use these to solve the integrals that come up in each
component, and find the following result for the Weyl spinor:
\begin{align}
	\Psi^{\Ad \Bd \Cd \Dd}_{(2)} &= \frac{1}{6r^{6}}\Bigg[u_{C}^{(\Ad} u_{B}^{\Dd)} r^{\dot{B}B }r^{\dot{C}C} -u_{B}^{(\Cd} u_{A}^{\Dd)} r^{\dot{A}A} r^{\dot{B}B}-u_{C}^{(\Bd} u_{A}^{\Dd)} r^{\dot{A}A}r^{\dot{C}C}\notag\\
	&-\left(u_{D}^{\Cd}\left(u_{A}^{\Bd} r^{\dot{A}A}+u_{B}^{\Ad} r^{\dot{B}B}\right)+u_{D}^{\Bd}\left(u_{A}^{\Cd} r^{\dot{A}A}+u_{C}^{\Ad} r^{\dot{C}C}\right)+u_{D}^{\Ad}\left(u_{B}^{\Cd} r^{\dot{B}B}+u_{C}^{\Bd} r^{\dot{C}C}\right)\right) r^{\dot{D}D}\Bigg],
\label{Weylres}
\end{align}
where $r_{\Bd B}$ is defined as in eq.~(\ref{rBB}). Remarkably, the
various terms combine in such a way as to yield the combination
\begin{equation}
  \Psi^{(n)}_{\Ad\Bd\Cd\Dd} = -2\frac{\Phi^{(2)}_{(\Ad\Bd}
    \Phi^{(2)}_{\Cd\Dd)}}{\phi^{(2)}},
  \label{Weylres2}
\end{equation}
in terms of the electromagnetic spinor of eq.~(\ref{phiABres2}). Thus,
a Weyl double copy does indeed hold for the integrated Weyl spinor,
even in the presence of a non-trivial source. We can also consider the
mixed index spinor, for which the integrated result turns out to be
\begin{align}
  	\Phi^{(2)}_{A B \dot{C} \dot{D}} &= -\frac{2}{r^6}\left(u_{\dot{A}}^B r^{A \dot{A}}+u_{\dot{B}}^A r^{B \dot{B}}\right)\left(u_C^{\dot{D}} r^{C \dot{C}}+u_C^{\dot{C}} r^{C \dot{D}}\right),
        \label{Mixedres}        
\end{align}
which reduces to the form
\begin{equation}
  \Phi^{(2)}_{AB\Cd\Dd}=-2\frac{\Phi^{(2)}_{AB}\bar{\Phi}^{(2)}_{\Cd\Dd}}
      {\phi^{(2)}},
  \label{Mixedres2}
\end{equation}
in terms of the electromagnetic (conjugate) spinors and scalar field
of sections~\ref{sec:scalarRN} and~\ref{sec:EM}. Thus, we also find a
double copy formula for the mixed index spinor, and we may immediately
compare this with the conjectured form of
refs.~\cite{Easson:2021asd,Easson:2022zoh}, given here in
eqs.~(\ref{Mantonmixed}). We indeed find agreement, which confirms the
conjecture for the special case of the momentum-space source
distribution of eq.~(\ref{JRN}). In order to go beyond this, we can in
principle repeat the arguments of this section for different
sources. In the following section, however, we see that an approach
exists, based directly on the tensorial double copy ideas of
refs.~\cite{Monteiro:2021ztt}.

\section{Sourced Weyl double copy from tensor methods}
\label{sec:tensor}

In the previous section, we have obtained a Weyl double copy formula
for the Reissner-Nordstrom black hole, which has a particular source
profile for the associated scalar field, as shown in
eq.~(\ref{rhoSKN}). Our results (including for the mixed index Ricci
spinor) agree with the more generally applicable formula conjectured
in refs.~\cite{Easson:2021asd,Easson:2022zoh}, and thus we now look at
generalising our conclusions to include sources which have an
arbitrary inverse power of the radial coordinate $r$. To do so, we
will introduce an alternative argument that complements the spinorial
approach of the previous sections, and recovers our former results as
a special case. Our starting point is a tensorial formula for the
conventional Weyl double copy, obtained in
ref.~\cite{Monteiro:2021ztt}:
\begin{equation}
W^{\mu\nu\rho\sigma} = \frac{1}{\phi}\cl{P}^{\mu\nu\rho\sigma}_{\tau \lambda \eta \omega}F^{\tau \lambda}F^{\eta \omega}.
\label{TensorDC}
\end{equation}
The Weyl tensor appears on the left-hand side, and this is then
expressed as a certain product of electromagnetic field-strength
tensors, contracted with a projector $\mathcal{P}$. The latter appears
in the definition of the Weyl tensor in terms of the Riemann curvature
tensor:
\[
	W^{\mu\nu\rho\sigma} = \cl{P}^{\mu\nu\rho\sigma}_{\tau \lambda \eta \omega}R^{\tau \lambda \eta \omega},
\label{Wproject}
        \]
and we thus see that the job of the projector is to remove traces from
the latter. It is given by
\[\label{projector}
\mathcal{P}_{\tau \lambda \eta \omega}^{\mu \nu \rho \sigma}=\delta_\tau^\mu \delta_\lambda^\nu \delta_\eta^\rho \delta_\omega^\sigma+\frac{1}{2} g_{\tau \eta} \delta_\lambda^{[\mu} g^{\nu][\rho} \delta_\omega^{\sigma]}+\frac{1}{6} g_{\tau \eta} g_{\lambda \omega} g^{\mu[\rho} g^{\sigma] \nu},
\]
where $g^{\mu\nu}$ is the full metric which, given that we are
considering solutions of the linearised Einstein equation, can be
replaced with the Minkowski metric $\eta^{\mu\nu}$. For the sourced
double copy, and based on eq.~(\ref{MantonDC}), it is natural to
generalise eq.~(\ref{TensorDC}) to 
\[
W^{\mu\nu\rho\sigma} = \sum_{n=1}^m\frac{f(n)}{\phi^{(n)}}\cl{P}^{\mu\nu\rho\sigma}_{\tau \lambda \eta \omega}F^{(n),\tau \lambda}F^{(n),\eta \omega}.
\label{tensorn}
\]
That is, we conjecture that the Weyl tensor can be written as a sum of
terms similar to eq.~(\ref{TensorDC}), and where we also allow for
some function $f(n)$ that introduces a potential normalisation
constant that can be different for each term. For each term, we will
consider a scalar field of the form
\begin{equation}
  \phi^{(n)} = \frac{1}{r^n},
  \label{phindef}
\end{equation}
where $r^2 = x^2 - (u\cdot x)^2$ the timelike vector associated with
the source. For static sources with $u^\mu=(1,\vec{0})$, $r$ then
reduces to the spatial radial coordinate. Each field in
eq.~(\ref{phindef}) satisfies an equation of motion
\[
\pd^2\phi^{(n)} = \frac{n(n-1)}{r^{2+n}} = \rho^{(n)}(r),
\]
with $n = 1$ being the vacuum case consistent with the standard Weyl
double copy. For the case of general $n$, we can use eqs.~(\ref{Fphi},
\ref{hphi}, \ref{Rphi}) to verify that the conjectured tensorial
double copy of eq.~(\ref{tensorn}) indeed holds. First, from
eq.~(\ref{Fphi}), the electromagnetic field strength tensor associated
with the scalar field of eq.~(\ref{phindef}) is
\[
  \label{sourcedF}
F_{\mu\nu}^{(n)} = n\frac{u_\mu r_\nu - u_\nu r_\mu}{r^{2+n}},
\]
with associated equation of motion
\[
\pd^\nu F_{\mu\nu}^{(n)} = \frac{n(n-1)}{r^{2+n}}u_\mu = J_\mu^{(n)} = \rho^{(n)}(r)u_\mu.
\]
Again we see that the vacuum solution is given by $n = 1$, which acts
as a cross-check of this expression. Furthermore, the source term has
the form expected from eq.~(\ref{rhoSdef}). Finally, one may associate
a gravity solution with the scalar field of eq.~(\ref{phindef}), where
the Riemann tensor will have the form of eq.~(\ref{Rphi}). One may
then write this in terms of the projector in eq. \eqref{projector},
which gives
\[
R_{(n)}^{\mu\nu\rho\sigma} = \cl{P}^{\mu\nu\rho\sigma}_{\tau \lambda \eta \omega}\pd^{[\tau}u^{\lambda]}\pd^{[\eta}u^{\omega]}\phi^{(n)}(x) - \left[\left(u^\mu u^{[\sigma}\eta^{\rho]\nu} - u^\nu u^{[\sigma}\eta^{\rho]\mu}\right)+\frac13\left(\eta^{\mu\sigma}\eta^{\nu\rho} - \eta^{\mu\rho}\eta^{\nu\sigma}\right)\right]\pd^2\phi^{(n)}(r).
\label{Rproject}
\]
The use of this is that in operating with the projector to obtain the
Weyl tensor, as given in eq.~(\ref{Wproject}), the second term on the
right-hand side of eq.~(\ref{Rproject}) vanishes. Thus, the Weyl
tensor is simply given by
\[
W_{(n)}^{\mu\nu\rho\sigma} &= \cl{P}^{\mu\nu\rho\sigma}_{\tau \lambda \eta \omega}\pd^{[\tau}u^{\lambda]}\pd^{[\eta}u^{\omega]}\phi^{(n)}(r)\\
&= \cl{P}^{\mu\nu\rho\sigma}_{\tau \lambda \eta \omega}u^{[\lambda}u^{[\omega}\pd^{\tau]}\pd^{\eta]}\phi^{(n)}(r),
\]
where the derivatives are understood as acting on all factors to the
right. We can relate the double derivative of the scalar to a
combination of single derivatives:
\[
    \pd_{\mu}\pd_{\nu}\phi^{(n)} = \frac{(2+n)}{n}\frac{\pd_\mu \phi^{(n)}\pd_\nu\phi^{(n)}}{\phi^{(n)}} - nr^{2n-2}(\phi^{(n)})^3(\eta_{\mu\nu} - u_\mu u_\nu).
\]
The second term is projected away when forming the Weyl tensor, and
thus we finally arrive at the result:
\[
    W^{\mu\nu\rho\sigma} &= \sum_n\frac{4(2+n)}{n}\cl{P}^{\mu\nu\rho\sigma}_{\tau \lambda \eta \omega}\frac{\pd^{[\tau}u^{\lambda]}\phi^{(n)}\pd^{[\eta}u^{\omega]}\phi^{(n)}}{\phi^{(n)}}\\
    &= \sum_n\frac{(2+n)}{n}\cl{P}^{\mu\nu\rho\sigma}_{\tau \lambda \eta \omega}\frac{F^{(n),\tau\lambda}F^{(n),\eta\omega}}{\phi^{(n)}}.
    \]
In other words, we confirm the sourced tensorial Weyl double copy of
eq.~(\ref{tensorn}), where the function entering each term is found to be
\begin{equation}
  f(n)=\frac{2+n}{n}.
  \label{fnres}
\end{equation}
Let us now connect this result with the spinorial analysis of the
previous section. Computing the Maxwell spinor for a given $n$ from
eq. \eqref{sourcedF} we find
\[
	\Phi_{AB}^{(n)} = -n\frac{u_{(A}^{~~\dot{B}}r_{B)\dot{B}}}{r^{2+n}}.
\]
Similarly, the Weyl spinor is found to be 
\[
	\Psi^{A B C D}_{(n)} &= \frac{1}{24}\frac{n(2+n)}{r^{4+n}}\Bigg[u_{\dot{C}}^{(A} u_{\dot{B}}^{D)} r^{B \dot{B}}r^{C \dot{C}} -u_{\dot{B}}^{(C} u_{\dot{A}}^{D)} r^{A \dot{A}} r^{B \dot{B}}-u_{\dot{C}}^{(B} u_{\dot{A}}^{D)} r^{A \dot{A}}r^{C \dot{C}}\\
	&-\left(u_{\dot{D}}^C\left(u_{\dot{A}}^B r^{A \dot{A}}+u_{\dot{B}}^A r^{B \dot{B}}\right)+u_{\dot{D}}^B\left(u_{\dot{A}}^C r^{A \dot{A}}+u_{\dot{C}}^A r^{C \dot{C}}\right)+u_{\dot{D}}^A\left(u_{\dot{B}}^C r^{B \dot{B}}+u_{\dot{C}}^B r^{C \dot{C}}\right)\right) r^{D \dot{D}}\Bigg]\\
	&= -\frac{2+n}{n}\frac{\varphi_n^{(AB}\varphi_n^{CD)}}{\phi^{(n)}},
\label{Weyln}
        \]
and for the mixed index case
\[
	\Phi^{A B \dot{C} \dot{D}}_{(n)} &= -\frac{1}{4} n(2+n) r^{-4-n}\left(u_{\dot{A}}^B r^{A \dot{A}}+u_{\dot{B}}^A r^{B \dot{B}}\right)\left(u_C^{\dot{D}} r^{C \dot{C}}+u_C^{\dot{C}} r^{C \dot{D}}\right)\\
	&= -\frac{2+n}{n}\frac{\varphi_n^{AB}\bar{\varphi}_n^{\dot{C}\dot{D}}}{\phi^{(n)}}.
        \label{mixedn}
\]
We see then that these spinors agree with the ones derived in the
previous section for the case of $n = 2$, up to overall
constants. However, the results are much broader than this, in that
eqs.~(\ref{Weyln}, \ref{mixedn}) provide a generalisation of the
spinorial version of the sourced Weyl double copy for scalar sources
involving arbitrary powers of the inverse radial coordinate. We thus
fully confirm the conjectures presented in ref.~\cite{Easson:2021asd}
for such solutions, but also go beyond the results in showing that the
Ricci spinor has the general double-copy form
\begin{equation}
  \Phi^{(n)}_{AB\Cd\Dd}
  =\sum_{n=1}^m \frac{1}{\phi^{(n)}}\Phi^{(n)}_{AB}\Phi^{(n)}_{\Cd\Dd},
    \label{RicciDC}
\end{equation}
where normalisation constants associated with each term have been
absorbed in the scalar field. This is the mixed-index equivalent of
the generalised Weyl double copy formula of ref.~(\ref{MantonDC}).

\section{Spinning Sources and the Weyl Double Copy}\label{sec:spin}
So far, we have considered the Weyl double copy for scalar sources,
which are spherically symmetric. However, our results can be
generalised beyond this to include other cases considered in
refs.~\cite{Easson:2021asd,Easson:2022zoh}, including the Kerr-Newman
black hole. The latter is a non-spherically symmetric charged black
hole solution, with non-zero intrinsic angular momentum. Its Weyl
double copy is straightforwardly inherited from the results of the
previous section, due to the existence of the well-known
\emph{Newman-Janis algorithm} \cite{Newman:1965tw}, which allows spinning gravity solutions
to be obtained from their non-spinning counterparts via a complex
translation. A similar procedure can be applied to the scalar field
and electromagnetic solutions, acting directly in the latter case on
the Maxwell spinor $\Phi_{AB}$ and its conjugate. Starting with the
electromagnetic and scalar fields corresponding to the
Reissner-Nordstrom black hole, we may then construct Weyl and Ricci
spinors using eqs.~(\ref{MantonDC}, \ref{Mantonmixed}), and verify
that they reproduce the known results for the Kerr-Newman black hole.

In more detail, the Newman-Janis prescription tells us to complexify
the radial coordinate in some particular way, and then to take $x^\mu
\rightarrow x^\mu + ia^\mu$, where $a^\mu$ is a vector describing the
angular momentum per unit mass. The Newman-Janis algorithm is usually
applied at the level of the metric, in the case of gravity, or the
gauge field in the case of electromagnetism. One prescription
in these cases is to complexify the radial coordinate as \cite{deUrreta:2015nla}
\[
r^n \rightarrow \frac{[\Re(r)]^{n+2}}{|r|^2},
\]
before performing the shift $x^\mu \rightarrow x^\mu + ia^\mu$. 
We are interested in scalar, gauge and gravitational potentials of the form $1/r^n$, complexified as above. In the scalar case, this is simply
\[
\phi^{(n)} = \frac{[\Re(r)]^{2-n}}{|r|^2},
\]
while for gauge fields, we have
\[
A_\mu^{(n)} = u_\mu\frac{[\Re(r)]^{2-n}}{|r|^2}.
\]
We can use this to compute the relevant field-strengths. For $n = 1$, we have
\[
F^{(1)}_{\mu\nu} = 4\frac{u_{[\mu}\hat{r}_{\nu]}}{r^2},
\]
while for $n = 2$ it is given by
\[
F^{(2)}_{\mu\nu} = 4\frac{u_{[\mu}\hat{r}_{\nu]}}{r|r|^2},
\] 
where we have expressed the position vector in spherical polar coordinates as
\[
r_\nu = r\hat{r}_\nu = r(0,\sin\theta\cos\phi,\sin\theta\sin\phi,\cos\theta).
\]
From these field-strengths, we can compute the Maxwell spinors, and then apply the complex translation to find the spinning solutions. For the $n = 1$ case, we find the Maxwell spinor is given by
\[
\Phi^{(1)}_{AB} = \frac{1}{(\tilde{r}+ia\cos\theta)^2}o_{(A}\iota_{B)}.
\]
while for $n = 2$, we find
\[
\Phi^{(2)}_{AB} = \frac{1}{(\tilde{r}^2+a^2\cos^2\theta)(\tilde{r}+ia\cos\theta)}o_{(A}\iota_{B)}.
\]
where we have chosen $a^\mu = (0,0,0,a)$ and $\tilde{r}$ is the radial coordinate in oblate-spheroidal coordinates, defined by
\[
  x+iy = \tilde{r}e^{i\theta}\sin\theta,~~~~~z = \tilde{r}\cos\theta.
\]
The radial coordinate $\tilde{r}$ is now implicitly a solution to the equation
\[\label{reqn}
	\frac{x^2+y^2}{\tilde{r}^2+a^2}+\frac{z^2}{\tilde{r}^2}=1.
\]
A particular solution to the equation \eqref{reqn} is given by
\[
\tilde{r} = \frac{1}{\sqrt{2}}\sqrt{r^2-a^2 + \sqrt{4a^2\cos^2\theta + (r^2-a^2)^2}},
\]
which has been chosen such that $\lim_{a\rightarrow 0}\tilde{r} = r$.
We note that in these coordinates, the basis spinors are given by
\[
o^A=\binom{\cos (\theta / 2)}{e^{i \phi} \sin (\theta / 2)}, \quad \imath^A=\binom{-\sin \left(\frac{\theta}{2}\right)}{e^{i \phi} \cos (\theta / 2)},
\]
which satisfy $o_A\iota^A = 1$ and the relations
\[
u^{A\Ad} &= o^A\bar{o}^{\Ad} + \iota^A\bar{\iota}^{\Ad}\\
\hat{r}^{A\Ad} &= o^A\bar{o}^{\Ad} - \iota^A\bar{\iota}^{\Ad}.
\]
We now have all of the ingredients we need to compute the gravitational spinors via the double copy, defined as before. We find that the Weyl spinor for the $n = 1$ case is given by
\[
\Psi^{ABCD}_{(1)} &= -3\frac{\Phi^{(AB}_{(1)}\Phi^{CD)}_{(1)}}{\phi_{(1)}}\\
&= -3\frac{GM}{(\tilde{r}+ia\cos\theta)^3}o^{(A}\iota^{B}o^{C}\iota^{D)},
\label{WKN1}
\] 
while for $n = 2$ it is
\[
  \Psi^{ABCD}_{(2)} &= -
  2\frac{\Phi^{(AB}_{(2)}\Phi^{CD)}_{(2)}}{\phi_{(2)}}\\ &=
  -
  2\frac{1}{(\tilde{r}^2+a^2\cos^2\theta)(\tilde{r}+ia\cos\theta)^2}o^{(A}\iota^{B}o^{C}\iota^{D)}.
  \label{WKN2}
\]
The sum of eqs.~(\ref{WKN1}, \ref{WKN2}) indeed reproduces the known
Weyl spinor for the Kerr-Newman solution, whilst also confirming the
form of eq.~(\ref{Mantonmixed}), as conjectured in
ref.~\cite{Easson:2022zoh}. We can also construct the Ricci spinor via
eq.~(\ref{Mantonmixed}), finding
\begin{equation}
  \Phi_{AB\Cd\Dd}=\frac{1}{(r^2+a^2\cos^2\theta)}o_{(A}\iota_{B)}\bar{o}_{(\Cd}
  \bar{\iota}_{\Dd)},
  \label{RicciKN}
\end{equation}
which again agrees with the known result.

As discussed in section~\ref{sec:review}, an alternative double-copy
form for the Ricci spinor is presented in ref.~\cite{Easson:2022zoh},
namely that of eq.~(\ref{Mantonmixed2}), and which is argued to be
more generally applicable than eq.~(\ref{Mantonmixed}), in that it
applies to solutions that are not asymptotically flat. In fact, it is
possible to give a general argument for this alternative formula, as
follows. First, one may consider the tensorial equations of motion for
Einstein-Maxwell theory: 
\[
R_{\mu\nu} - \frac12 g_{\mu\nu}R &= F_{\mu\alpha}F_\nu^{~\alpha} - \frac14 g_{\mu\nu}F_{\alpha\beta}F^{\alpha\beta}\\
&= \frac12F_{\mu\alpha}^+F_{\nu}^{~\alpha-},
\label{EMEOM}
\]
where in the second line we have written the right-hand side
explicitly in terms of (anti-)self-dual field strengths. Taking the
trace of eq.~(\ref{EMEOM}) yields $R=0$, due to the tracelessness of
the electromagnetic field strength tensor. This in turn simplifies the
calculation of the Ricci spinor, as
\[
  R_{\mu\nu}\sigma^{\mu}_{A\Ad}\sigma^{\nu}_{B\Bd} = \frac12F_{\mu\alpha}^+F_{\nu}^{~\alpha-}\sigma^{\mu}_{A\Ad}\sigma^{\nu}_{B\Bd}.
\]
The spinorial translations of the (anti-)self-dual field strength
tensors are simply given by
\begin{equation}
  \frac12 F^+_{A\Ad B\Bd}= \Phi_{AB}\epsilon_{\Ad\Bd},\quad
  \frac12 F^-_{A\Ad B\Bd}= \bar{\Phi}_{\Ad \Bd}\epsilon_{AB}.
  \label{F+-spinor}
\end{equation}
One then finds that the Ricci spinor is given by
\[
\Phi_{AB\Ad\Bd} = 2\Phi_{AB}\bar{\Phi}_{\Ad\Bd}.
\label{DCmixedres}
\]
Note that this is not yet a double-copy formula:
eq.~(\ref{DCmixedres}) relates the Ricci spinor of an Einstein-Maxwell
solution to the electromagnetic spinors of the same solution {\it in
  the same theory}. This is different to the double copy, which
relates quantities in a gravitational theory (possibly with extra
fields) to quantities in a {\it different} ``single-copy''
theory. However, it is often the case that the purely electromagnetic
part of an Einstein-Maxwell solution, obtained by setting the
gravitational constant $G_N\rightarrow 0$, corresponds with the
single-copy gauge field. This is certainly true for the Kerr-Newman
black hole~\cite{Monteiro:2014cda}. For such solutions, we may
therefore also interpret eq.~(\ref{DCmixedres}) as a genuine
double-copy formula, relating a gravitational quantity on the left, to
single-copy quantities on the right.

It may seem that eq.~(\ref{DCmixedres}) contradicts the double copy
formula of eq.~(\ref{RicciDC}). However, the fact that $\Phi_{AB}$
above is associated with the vacuum part of the single-copy
(electromagnetic) solution means that it corresponds to the $n=1$ term in
eq.~(\ref{MantonDC}). Equation~(\ref{RicciDC}) instead involves the
$n=2$ spinors, and it is indeed the case that
\[
R_{AB\Ad\Bd} = 2\Phi^{(1)}_{AB}\bar{\Phi}^{(1)}_{\Ad\Bd} = 2\frac{\Phi^{(2)}_{AB}\bar{\Phi}^{(2)}_{\Ad\Bd}}{\phi^{(2)}}.
\]
We thus obtain the two double copy prescriptions for the Ricci spinor
(where applicable) of ref.~\cite{Easson:2022zoh}. 

\section{Conclusion}
\label{sec:conclude}

In this paper, we have addressed the systematic derivation of Weyl
double copy formulae for sourced (non-vacuum) classical solutions. In
such cases, the spinorial translation of the Riemann curvature tensor
gives rise to a number of different spinor quantities, including a
mixed-index Ricci spinor in addition to the Weyl
spinor. References~\cite{Easson:2021asd,Easson:2022zoh} conjectured a
generalised Weyl double copy formula for the Weyl tensor, involving a sum
of double-copy terms, as shown here in eqs.~(\ref{MantonDC}). In each
term, the gravity, gauge and scalar theories are linked by having
physically similar source terms, and we here confirm the conjecture by
focusing on sources that can be expanded in arbitrary powers of the
inverse radial coordinate. We demonstrate the double copy using two
methods: (i) a spinorial approach inspired by the twistor double copy
approach of
refs.~\cite{White:2020sfn,Chacon:2021wbr,Chacon:2021lox,Chacon:2021hfe,Luna:2022dxo,CarrilloGonzalez:2022ggn,Armstrong-Williams:2023ssz};
(ii) a tensorial approach inspired by ref.~\cite{Monteiro:2021ztt},
and which reproduces the same results as the spinorial approach where
our analyses overlap. Furthermore, we present a generalised double
copy formula for the Ricci spinor, involving a similar sum of terms
with physically similar sources. This generalises the results
presented in refs.~\cite{Easson:2021asd,Easson:2022zoh}, which were
limited to Einstein-Maxwell theory, for which only two terms are
required in the sum. Going beyond spherically symmetric sources, we
obtained the Kerr-Newman black hole by taking a Newman-Janis shift of
our previous results. This in turn clarified the fact that the two
alternative double copy formulae for the Ricci spinor given in
ref.~\cite{Easson:2022zoh} are indeed equivalent.

There are a number of directions for further work. One interesting
avenue would be to explore whether non-vacuum double copies may have
applications to astrophysics / cosmology, or to the generation of new
gravitational solutions. Also, our results are for linearised gauge
/ gravity theories only, and one might think of trying to generalise
the Weyl double copy -- or alternative classical double copy
approaches -- to non-linear level. We hope that our results prove
useful in thinking about these, and other, issues.

\section*{Acknowledgments}

We are very grateful to Tucker Manton for numerous useful
discussions. This work has been supported by the UK Science and
Technology Facilities Council (STFC) Consolidated Grant ST/P000754/1
``String theory, gauge theory and duality''. KAW is supported by a
studentship from the UK Engineering and Physical Sciences Research
Council (EPSRC).

\bibliography{refs}

\providecommand{\href}[2]{#2}\begingroup\raggedright\begin{thebibliography}{10}

\bibitem{Bern:2010ue}
Z.~Bern, J.~J.~M. Carrasco, and H.~Johansson, ``{Perturbative Quantum Gravity
  as a Double Copy of Gauge Theory},'' {\em Phys.Rev.Lett.} {\bf 105} (2010)
  061602, \href{http://www.arXiv.org/abs/1004.0476}{{\tt 1004.0476}}.

\bibitem{Bern:2010yg}
Z.~Bern, T.~Dennen, Y.-t. Huang, and M.~Kiermaier, ``{Gravity as the Square of
  Gauge Theory},'' {\em Phys.Rev.} {\bf D82} (2010) 065003,
  \href{http://www.arXiv.org/abs/1004.0693}{{\tt 1004.0693}}.

\bibitem{Kawai:1985xq}
H.~Kawai, D.~Lewellen, and S.~Tye, ``{A Relation Between Tree Amplitudes of
  Closed and Open Strings},'' {\em Nucl.Phys.} {\bf B269} (1986)
1.
%%CITATION = NUPHA,B269,1;%%.

\bibitem{Monteiro:2014cda}
R.~Monteiro, D.~O'Connell, and C.~D. White, ``{Black holes and the double
  copy},'' {\em JHEP} {\bf 1412} (2014) 056,
\href{http://www.arXiv.org/abs/1410.0239}{{\tt 1410.0239}}.
%%CITATION = ARXIV:1410.0239;%%.

\bibitem{Luna:2015paa}
A.~Luna, R.~Monteiro, D.~O'Connell, and C.~D. White, ``{The classical double
  copy for Taub-NUT spacetime},'' {\em Phys. Lett.} {\bf B750} (2015) 272--277,
\href{http://www.arXiv.org/abs/1507.01869}{{\tt 1507.01869}}.
%%CITATION = ARXIV:1507.01869;%%.

\bibitem{Ridgway:2015fdl}
A.~K. Ridgway and M.~B. Wise, ``{Static Spherically Symmetric Kerr-Schild
  Metrics and Implications for the Classical Double Copy},'' {\em Phys. Rev.}
  {\bf D94} (2016), no.~4, 044023,
\href{http://www.arXiv.org/abs/1512.02243}{{\tt 1512.02243}}.
%%CITATION = ARXIV:1512.02243;%%.

\bibitem{Bahjat-Abbas:2017htu}
N.~Bahjat-Abbas, A.~Luna, and C.~D. White, ``{The Kerr-Schild double copy in
  curved spacetime},'' {\em JHEP} {\bf 12} (2017) 004,
\href{http://www.arXiv.org/abs/1710.01953}{{\tt 1710.01953}}.
%%CITATION = ARXIV:1710.01953;%%.

\bibitem{Carrillo-Gonzalez:2017iyj}
M.~Carrillo-González, R.~Penco, and M.~Trodden, ``{The classical double copy
  in maximally symmetric spacetimes},'' {\em JHEP} {\bf 04} (2018) 028,
\href{http://www.arXiv.org/abs/1711.01296}{{\tt 1711.01296}}.
%%CITATION = ARXIV:1711.01296;%%.

\bibitem{CarrilloGonzalez:2019gof}
M.~Carrillo~González, B.~Melcher, K.~Ratliff, S.~Watson, and C.~D. White,
  ``{The classical double copy in three spacetime dimensions},'' {\em JHEP}
  {\bf 07} (2019) 167,
\href{http://www.arXiv.org/abs/1904.11001}{{\tt 1904.11001}}.
%%CITATION = ARXIV:1904.11001;%%.

\bibitem{Bah:2019sda}
I.~Bah, R.~Dempsey, and P.~Weck, ``{Kerr-Schild Double Copy and Complex
  Worldlines},''
\href{http://www.arXiv.org/abs/1910.04197}{{\tt 1910.04197}}.
%%CITATION = ARXIV:1910.04197;%%.

\bibitem{Alkac:2021seh}
G.~Alkac, M.~K. Gumus, and M.~A. Olpak, ``{The Kerr-Schild Double Copy of the
  Coulomb Solution in Three Dimensions},''
  \href{http://www.arXiv.org/abs/2105.11550}{{\tt 2105.11550}}.

\bibitem{Alkac:2022tvc}
G.~Alkac, M.~K. Gumus, and M.~A. Olpak, ``{Generalized black holes in 3D
  Kerr-Schild double copy},'' {\em Phys. Rev. D} {\bf 106} (2022), no.~2,
  026013, \href{http://www.arXiv.org/abs/2205.08503}{{\tt 2205.08503}}.

\bibitem{Luna:2018dpt}
A.~Luna, R.~Monteiro, I.~Nicholson, and D.~O'Connell, ``{Type D Spacetimes and
  the Weyl Double Copy},'' {\em Class. Quant. Grav.} {\bf 36} (2019) 065003,
\href{http://www.arXiv.org/abs/1810.08183}{{\tt 1810.08183}}.
%%CITATION = ARXIV:1810.08183;%%.

\bibitem{Sabharwal:2019ngs}
S.~Sabharwal and J.~W. Dalhuisen, ``{Anti-Self-Dual Spacetimes, Gravitational
  Instantons and Knotted Zeros of the Weyl Tensor},'' {\em JHEP} {\bf 07}
  (2019) 004, \href{http://www.arXiv.org/abs/1904.06030}{{\tt 1904.06030}}.

\bibitem{Alawadhi:2020jrv}
R.~Alawadhi, D.~S. Berman, and B.~Spence, ``{Weyl doubling},'' {\em JHEP} {\bf
  09} (2020) 127, \href{http://www.arXiv.org/abs/2007.03264}{{\tt 2007.03264}}.

\bibitem{Godazgar:2020zbv}
H.~Godazgar, M.~Godazgar, R.~Monteiro, D.~Peinador~Veiga, and C.~N. Pope,
  ``{Weyl Double Copy for Gravitational Waves},'' {\em Phys. Rev. Lett.} {\bf
  126} (2021), no.~10, 101103, \href{http://www.arXiv.org/abs/2010.02925}{{\tt
  2010.02925}}.

\bibitem{White:2020sfn}
C.~D. White, ``{Twistorial Foundation for the Classical Double Copy},'' {\em
  Phys. Rev. Lett.} {\bf 126} (2021), no.~6, 061602,
  \href{http://www.arXiv.org/abs/2012.02479}{{\tt 2012.02479}}.

\bibitem{Chacon:2020fmr}
E.~Chac\'on, H.~Garc\'\i{}a-Compe\'an, A.~Luna, R.~Monteiro, and C.~D. White,
  ``{New heavenly double copies},'' {\em JHEP} {\bf 03} (2021) 247,
  \href{http://www.arXiv.org/abs/2008.09603}{{\tt 2008.09603}}.

\bibitem{Chacon:2021wbr}
E.~Chac\'on, S.~Nagy, and C.~D. White, ``{The Weyl double copy from twistor
  space},'' {\em JHEP} {\bf 05} (2021) 2239,
  \href{http://www.arXiv.org/abs/2103.16441}{{\tt 2103.16441}}.

\bibitem{Chacon:2021hfe}
E.~Chac\'on, A.~Luna, and C.~D. White, ``{Double copy of the multipole
  expansion},'' {\em Phys. Rev. D} {\bf 106} (2022), no.~8, 086020,
  \href{http://www.arXiv.org/abs/2108.07702}{{\tt 2108.07702}}.

\bibitem{Chacon:2021lox}
E.~Chac\'on, S.~Nagy, and C.~D. White, ``{Alternative formulations of the
  twistor double copy},'' {\em JHEP} {\bf 03} (2022) 180,
  \href{http://www.arXiv.org/abs/2112.06764}{{\tt 2112.06764}}.

\bibitem{Dempsey:2022sls}
R.~Dempsey and P.~Weck, ``{Compactifying the Kerr-Schild Double Copy},''
  \href{http://www.arXiv.org/abs/2211.14327}{{\tt 2211.14327}}.

\bibitem{Easson:2022zoh}
D.~A. Easson, T.~Manton, and A.~Svesko, ``{Einstein-Maxwell theory and the Weyl
  double copy},'' {\em Phys. Rev. D} {\bf 107} (2023), no.~4, 044063,
  \href{http://www.arXiv.org/abs/2210.16339}{{\tt 2210.16339}}.

\bibitem{Chawla:2022ogv}
S.~Chawla and C.~Keeler, ``{Aligned Fields Double Copy to Kerr-NUT-(A)dS},''
  \href{http://www.arXiv.org/abs/2209.09275}{{\tt 2209.09275}}.

\bibitem{Han:2022mze}
S.~Han, ``{The Weyl double copy in vacuum spacetimes with a cosmological
  constant},'' {\em JHEP} {\bf 09} (2022) 238,
  \href{http://www.arXiv.org/abs/2205.08654}{{\tt 2205.08654}}.

\bibitem{Armstrong-Williams:2022apo}
K.~Armstrong-Williams, C.~D. White, and S.~Wikeley, ``{Non-perturbative aspects
  of the self-dual double copy},'' {\em JHEP} {\bf 08} (2022) 160,
  \href{http://www.arXiv.org/abs/2205.02136}{{\tt 2205.02136}}.

\bibitem{Han:2022ubu}
S.~Han, ``{Weyl double copy and massless free fields in curved spacetimes},''
  \href{http://www.arXiv.org/abs/2204.01907}{{\tt 2204.01907}}.

\bibitem{Elor:2020nqe}
G.~Elor, K.~Farnsworth, M.~L. Graesser, and G.~Herczeg, ``{The Newman-Penrose
  Map and the Classical Double Copy},''
  \href{http://www.arXiv.org/abs/2006.08630}{{\tt 2006.08630}}.

\bibitem{Farnsworth:2021wvs}
K.~Farnsworth, M.~L. Graesser, and G.~Herczeg, ``{Twistor Space Origins of the
  Newman-Penrose Map},'' \href{http://www.arXiv.org/abs/2104.09525}{{\tt
  2104.09525}}.

\bibitem{Anastasiou:2014qba}
A.~Anastasiou, L.~Borsten, M.~J. Duff, L.~J. Hughes, and S.~Nagy, ``{Yang-Mills
  origin of gravitational symmetries},'' {\em Phys. Rev. Lett.} {\bf 113}
  (2014), no.~23, 231606,
\href{http://www.arXiv.org/abs/1408.4434}{{\tt 1408.4434}}.
%%CITATION = ARXIV:1408.4434;%%.

\bibitem{LopesCardoso:2018xes}
G.~Lopes~Cardoso, G.~Inverso, S.~Nagy, and S.~Nampuri, ``{Comments on the
  double copy construction for gravitational theories},'' in {\em {17th
  Hellenic School and Workshops on Elementary Particle Physics and Gravity
  (CORFU2017) Corfu, Greece, September 2-28, 2017}}.
\newblock 2018.
\newblock
\href{http://www.arXiv.org/abs/1803.07670}{{\tt 1803.07670}}.
\newblock
%%CITATION = ARXIV:1803.07670;%%.

\bibitem{Anastasiou:2018rdx}
A.~Anastasiou, L.~Borsten, M.~J. Duff, S.~Nagy, and M.~Zoccali, ``{Gravity as
  Gauge Theory Squared: A Ghost Story},'' {\em Phys. Rev. Lett.} {\bf 121}
  (2018), no.~21, 211601,
\href{http://www.arXiv.org/abs/1807.02486}{{\tt 1807.02486}}.
%%CITATION = ARXIV:1807.02486;%%.

\bibitem{Luna:2020adi}
A.~Luna, S.~Nagy, and C.~White, ``{The convolutional double copy: a case study
  with a point},'' {\em JHEP} {\bf 09} (2020) 062,
  \href{http://www.arXiv.org/abs/2004.11254}{{\tt 2004.11254}}.

\bibitem{Borsten:2020xbt}
L.~Borsten and S.~Nagy, ``{The pure BRST Einstein-Hilbert Lagrangian from the
  double-copy to cubic order},'' {\em JHEP} {\bf 07} (2020) 093,
  \href{http://www.arXiv.org/abs/2004.14945}{{\tt 2004.14945}}.

\bibitem{Borsten:2020zgj}
L.~Borsten, B.~Jurco, H.~Kim, T.~Macrelli, C.~Saemann, and M.~Wolf,
  ``{Becchi-Rouet-Stora-Tyutin-Lagrangian Double Copy of Yang-Mills Theory},''
  {\em Phys. Rev. Lett.} {\bf 126} (2021), no.~19, 191601,
  \href{http://www.arXiv.org/abs/2007.13803}{{\tt 2007.13803}}.

\bibitem{Goldberger:2017frp}
W.~D. Goldberger, S.~G. Prabhu, and J.~O. Thompson, ``{Classical gluon and
  graviton radiation from the bi-adjoint scalar double copy},'' {\em Phys.
  Rev.} {\bf D96} (2017), no.~6, 065009,
\href{http://www.arXiv.org/abs/1705.09263}{{\tt 1705.09263}}.
%%CITATION = ARXIV:1705.09263;%%.

\bibitem{Goldberger:2017vcg}
W.~D. Goldberger and A.~K. Ridgway, ``{Bound states and the classical double
  copy},'' {\em Phys. Rev.} {\bf D97} (2018), no.~8, 085019,
\href{http://www.arXiv.org/abs/1711.09493}{{\tt 1711.09493}}.
%%CITATION = ARXIV:1711.09493;%%.

\bibitem{Goldberger:2017ogt}
W.~D. Goldberger, J.~Li, and S.~G. Prabhu, ``{Spinning particles, axion
  radiation, and the classical double copy},'' {\em Phys. Rev.} {\bf D97}
  (2018), no.~10, 105018,
\href{http://www.arXiv.org/abs/1712.09250}{{\tt 1712.09250}}.
%%CITATION = ARXIV:1712.09250;%%.

\bibitem{Goldberger:2019xef}
W.~D. Goldberger and J.~Li, ``{Strings, extended objects, and the classical
  double copy},''
\href{http://www.arXiv.org/abs/1912.01650}{{\tt 1912.01650}}.
%%CITATION = ARXIV:1912.01650;%%.

\bibitem{Goldberger:2016iau}
W.~D. Goldberger and A.~K. Ridgway, ``{Radiation and the classical double copy
  for color charges},'' {\em Phys. Rev.} {\bf D95} (2017), no.~12, 125010,
\href{http://www.arXiv.org/abs/1611.03493}{{\tt 1611.03493}}.
%%CITATION = ARXIV:1611.03493;%%.

\bibitem{Prabhu:2020avf}
S.~G. Prabhu, ``{The classical double copy in curved spacetimes: Perturbative
  Yang-Mills from the bi-adjoint scalar},''
  \href{http://www.arXiv.org/abs/2011.06588}{{\tt 2011.06588}}.

\bibitem{Luna:2016hge}
A.~Luna, R.~Monteiro, I.~Nicholson, A.~Ochirov, D.~O'Connell, N.~Westerberg,
  and C.~D. White, ``{Perturbative spacetimes from Yang-Mills theory},'' {\em
  JHEP} {\bf 04} (2017) 069,
\href{http://www.arXiv.org/abs/1611.07508}{{\tt 1611.07508}}.
%%CITATION = ARXIV:1611.07508;%%.

\bibitem{Luna:2017dtq}
A.~Luna, I.~Nicholson, D.~O'Connell, and C.~D. White, ``{Inelastic Black Hole
  Scattering from Charged Scalar Amplitudes},'' {\em JHEP} {\bf 03} (2018) 044,
\href{http://www.arXiv.org/abs/1711.03901}{{\tt 1711.03901}}.
%%CITATION = ARXIV:1711.03901;%%.

\bibitem{Cheung:2016prv}
C.~Cheung and C.-H. Shen, ``{Symmetry for Flavor-Kinematics Duality from an
  Action},'' {\em Phys. Rev. Lett.} {\bf 118} (2017), no.~12, 121601,
  \href{http://www.arXiv.org/abs/1612.00868}{{\tt 1612.00868}}.

\bibitem{Cheung:2021zvb}
C.~Cheung and J.~Mangan, ``{Covariant color-kinematics duality},'' {\em JHEP}
  {\bf 11} (2021) 069, \href{http://www.arXiv.org/abs/2108.02276}{{\tt
  2108.02276}}.

\bibitem{Cheung:2022vnd}
C.~Cheung, A.~Helset, and J.~Parra-Martinez, ``{Geometry-kinematics duality},''
  {\em Phys. Rev. D} {\bf 106} (2022), no.~4, 045016,
  \href{http://www.arXiv.org/abs/2202.06972}{{\tt 2202.06972}}.

\bibitem{Cheung:2022mix}
C.~Cheung, J.~Mangan, J.~Parra-Martinez, and N.~Shah, ``{Non-perturbative
  Double Copy in Flatland},'' {\em Phys. Rev. Lett.} {\bf 129} (2022), no.~22,
  221602, \href{http://www.arXiv.org/abs/2204.07130}{{\tt 2204.07130}}.

\bibitem{Chawla:2023bsu}
S.~Chawla and C.~Keeler, ``{Black hole horizons from the double copy},'' {\em
  Class. Quant. Grav.} {\bf 40} (2023), no.~22, 225004,
  \href{http://www.arXiv.org/abs/2306.02417}{{\tt 2306.02417}}.

\bibitem{Easson:2023dbk}
D.~A. Easson, G.~Herczeg, T.~Manton, and M.~Pezzelle, ``{Isometries and the
  double copy},'' {\em JHEP} {\bf 09} (2023) 162,
  \href{http://www.arXiv.org/abs/2306.13687}{{\tt 2306.13687}}.

\bibitem{Farnsworth:2023mff}
K.~Farnsworth, M.~L. Graesser, and G.~Herczeg, ``{Double Kerr-Schild spacetimes
  and the Newman-Penrose map},'' {\em JHEP} {\bf 10} (2023) 010,
  \href{http://www.arXiv.org/abs/2306.16445}{{\tt 2306.16445}}.

\bibitem{Borsten:2023paw}
L.~Borsten, B.~Jurco, H.~Kim, T.~Macrelli, C.~Saemann, and M.~Wolf,
  ``{Double-copying self-dual Yang-Mills theory to self-dual gravity on twistor
  space},'' {\em JHEP} {\bf 11} (2023) 172,
  \href{http://www.arXiv.org/abs/2307.10383}{{\tt 2307.10383}}.

\bibitem{Alkac:2023glx}
G.~Alkac, M.~K. Gumus, O.~Kasikci, M.~A. Olpak, and M.~Tek, ``{Regularized Weyl
  double copy},'' {\em Phys. Rev. D} {\bf 109} (2024), no.~8, 084047,
  \href{http://www.arXiv.org/abs/2310.06048}{{\tt 2310.06048}}.

\bibitem{He:2023iew}
J.-L. He and J.-H. Huang, ``{Cosmological horizons from classical double
  copy},'' {\em Phys. Lett. B} {\bf 851} (2024) 138579,
  \href{http://www.arXiv.org/abs/2312.00972}{{\tt 2312.00972}}.

\bibitem{Chawla:2024mse}
S.~Chawla, K.~Fransen, and C.~Keeler, ``{The Penrose limit of the Weyl double
  copy},'' \href{http://www.arXiv.org/abs/2406.14601}{{\tt 2406.14601}}.

\bibitem{Monteiro:2011pc}
R.~Monteiro and D.~O'Connell, ``{The Kinematic Algebra From the Self-Dual
  Sector},'' {\em JHEP} {\bf 1107} (2011) 007,
\href{http://www.arXiv.org/abs/1105.2565}{{\tt 1105.2565}}.
%%CITATION = ARXIV:1105.2565;%%.

\bibitem{Borsten:2021hua}
L.~Borsten, H.~Kim, B.~Jurco, T.~Macrelli, C.~Saemann, and M.~Wolf, ``{Double
  Copy from Homotopy Algebras},'' {\em Fortsch. Phys.} {\bf 69} (2021),
  no.~8-9, 2100075, \href{http://www.arXiv.org/abs/2102.11390}{{\tt
  2102.11390}}.

\bibitem{Alawadhi:2019urr}
R.~Alawadhi, D.~S. Berman, B.~Spence, and D.~Peinador~Veiga, ``{S-duality and
  the double copy},'' {\em JHEP} {\bf 03} (2020) 059,
  \href{http://www.arXiv.org/abs/1911.06797}{{\tt 1911.06797}}.

\bibitem{Banerjee:2019saj}
A.~Banerjee, E.~Colgáin, J.~A. Rosabal, and H.~Yavartanoo, ``{Ehlers as EM
  duality in the double copy},''
\href{http://www.arXiv.org/abs/1912.02597}{{\tt 1912.02597}}.
%%CITATION = ARXIV:1912.02597;%%.

\bibitem{Huang:2019cja}
Y.-T. Huang, U.~Kol, and D.~O'Connell, ``{The Double Copy of Electric-Magnetic
  Duality},''
\href{http://www.arXiv.org/abs/1911.06318}{{\tt 1911.06318}}.
%%CITATION = ARXIV:1911.06318;%%.

\bibitem{Berman:2018hwd}
D.~S. Berman, E.~Chacón, A.~Luna, and C.~D. White, ``{The self-dual classical
  double copy, and the Eguchi-Hanson instanton},''
\href{http://www.arXiv.org/abs/1809.04063}{{\tt 1809.04063}}.
%%CITATION = ARXIV:1809.04063;%%.

\bibitem{Alfonsi:2020lub}
L.~Alfonsi, C.~D. White, and S.~Wikeley, ``{Topology and Wilson lines: global
  aspects of the double copy},'' {\em JHEP} {\bf 07} (2020) 091,
  \href{http://www.arXiv.org/abs/2004.07181}{{\tt 2004.07181}}.

\bibitem{Alawadhi:2021uie}
R.~Alawadhi, D.~S. Berman, C.~D. White, and S.~Wikeley, ``{The single copy of
  the gravitational holonomy},''
  \href{http://www.arXiv.org/abs/2107.01114}{{\tt 2107.01114}}.

\bibitem{White:2016jzc}
C.~D. White, ``{Exact solutions for the biadjoint scalar field},'' {\em Phys.
  Lett.} {\bf B763} (2016) 365--369,
\href{http://www.arXiv.org/abs/1606.04724}{{\tt 1606.04724}}.
%%CITATION = ARXIV:1606.04724;%%.

\bibitem{DeSmet:2017rve}
P.-J. De~Smet and C.~D. White, ``{Extended solutions for the biadjoint scalar
  field},'' {\em Phys. Lett.} {\bf B775} (2017) 163--167,
\href{http://www.arXiv.org/abs/1708.01103}{{\tt 1708.01103}}.
%%CITATION = ARXIV:1708.01103;%%.

\bibitem{Bahjat-Abbas:2018vgo}
N.~Bahjat-Abbas, R.~Stark-Muchão, and C.~D. White, ``{Biadjoint wires},'' {\em
  Phys. Lett.} {\bf B788} (2019) 274--279,
\href{http://www.arXiv.org/abs/1810.08118}{{\tt 1810.08118}}.
%%CITATION = ARXIV:1810.08118;%%.

\bibitem{Moynihan:2021rwh}
N.~Moynihan, ``{Massive Covariant Colour-Kinematics in 3D},''
  \href{http://www.arXiv.org/abs/2110.02209}{{\tt 2110.02209}}.

\bibitem{Borsten:2022vtg}
L.~Borsten, B.~Jurco, H.~Kim, T.~Macrelli, C.~Saemann, and M.~Wolf,
  ``{Kinematic Lie Algebras From Twistor Spaces},''
  \href{http://www.arXiv.org/abs/2211.13261}{{\tt 2211.13261}}.

\bibitem{Armstrong-Williams:2024icu}
K.~Armstrong-Williams, S.~Nagy, C.~D. White, and S.~Wikeley, ``{What can
  abelian gauge theories teach us about kinematic algebras?},''
  \href{http://www.arXiv.org/abs/2401.10750}{{\tt 2401.10750}}.

\bibitem{Borsten:2020bgv}
L.~Borsten, ``{Gravity as the square of gauge theory: a review},'' {\em Riv.
  Nuovo Cim.} {\bf 43} (2020), no.~3, 97--186.

\bibitem{Bern:2019prr}
Z.~Bern, J.~J. Carrasco, M.~Chiodaroli, H.~Johansson, and R.~Roiban, ``{The
  Duality Between Color and Kinematics and its Applications},''
\href{http://www.arXiv.org/abs/1909.01358}{{\tt 1909.01358}}.
%%CITATION = ARXIV:1909.01358;%%.

\bibitem{Adamo:2022dcm}
T.~Adamo, J.~J.~M. Carrasco, M.~Carrillo-Gonz\'alez, M.~Chiodaroli, H.~Elvang,
  H.~Johansson, D.~O'Connell, R.~Roiban, and O.~Schlotterer, ``{Snowmass White
  Paper: the Double Copy and its Applications},'' in {\em {2022 Snowmass Summer
  Study}}.
\newblock 4, 2022.
\newblock \href{http://www.arXiv.org/abs/2204.06547}{{\tt 2204.06547}}.

\bibitem{Bern:2022wqg}
Z.~Bern, J.~J. Carrasco, M.~Chiodaroli, H.~Johansson, and R.~Roiban, ``{Chapter
  2: An invitation to color-kinematics duality and the double copy},'' {\em J.
  Phys. A} {\bf 55} (2022), no.~44, 443003,
  \href{http://www.arXiv.org/abs/2203.13013}{{\tt 2203.13013}}.

\bibitem{White:2021gvv}
C.~D. White, ``{Double copy\textemdash{}from optics to quantum gravity:
  tutorial},'' {\em J. Opt. Soc. Am. B} {\bf 38} (2021), no.~11, 3319--3330,
  \href{http://www.arXiv.org/abs/2105.06809}{{\tt 2105.06809}}.

\bibitem{White:2024pve}
C.~D. White, {\em {The Classical Double Copy}}.
\newblock World Scientific, 5, 2024.

\bibitem{Luna:2022dxo}
A.~Luna, N.~Moynihan, and C.~D. White, ``{Why is the Weyl double copy local in
  position space?},'' {\em JHEP} {\bf 12} (2022) 046,
  \href{http://www.arXiv.org/abs/2208.08548}{{\tt 2208.08548}}.

\bibitem{Easson:2021asd}
D.~A. Easson, T.~Manton, and A.~Svesko, ``{Sources in the Weyl Double Copy},''
  {\em Phys. Rev. Lett.} {\bf 127} (2021), no.~27, 271101,
  \href{http://www.arXiv.org/abs/2110.02293}{{\tt 2110.02293}}.

\bibitem{CarrilloGonzalez:2022ggn}
M.~Carrillo~Gonz\'alez, W.~T. Emond, N.~Moynihan, J.~Rumbutis, and C.~D. White,
  ``{Mini-twistors and the Cotton double copy},'' {\em JHEP} {\bf 03} (2023)
  177, \href{http://www.arXiv.org/abs/2212.04783}{{\tt 2212.04783}}.

\bibitem{Armstrong-Williams:2023ssz}
K.~Armstrong-Williams and C.~D. White, ``{A spinorial double copy for $
  \mathcal{N} $ = 0 supergravity},'' {\em JHEP} {\bf 05} (2023) 047,
  \href{http://www.arXiv.org/abs/2303.04631}{{\tt 2303.04631}}.

\bibitem{Guevara:2021yud}
A.~Guevara, ``{Reconstructing Classical Spacetimes from the S-Matrix in Twistor
  Space},'' \href{http://www.arXiv.org/abs/2112.05111}{{\tt 2112.05111}}.

\bibitem{Penrose:1967wn}
R.~Penrose, ``{Twistor algebra},'' {\em J. Math. Phys.} {\bf 8} (1967) 345.

\bibitem{Penrose:1972ia}
R.~Penrose and M.~A.~H. MacCallum, ``{Twistor theory: An Approach to the
  quantization of fields and space-time},'' {\em Phys. Rept.} {\bf 6} (1972)
  241--316.

\bibitem{Eastwood:1981jy}
M.~G. Eastwood, R.~Penrose, and R.~O. Wells, ``{Cohomology and Massless
  Fields},'' {\em Commun. Math. Phys.} {\bf 78} (1981) 305--351.

\bibitem{Emond:2022uaf}
W.~T. Emond and N.~Moynihan, ``{Scattering Amplitudes and The Cotton Double
  Copy},'' \href{http://www.arXiv.org/abs/2202.10499}{{\tt 2202.10499}}.

\bibitem{Gonzalez:2022otg}
M.~C. Gonz\'alez, A.~Momeni, and J.~Rumbutis, ``{Cotton Double Copy for
  Gravitational Waves},'' \href{http://www.arXiv.org/abs/2202.10476}{{\tt
  2202.10476}}.

\bibitem{Moynihan:2020ejh}
N.~Moynihan, ``{Scattering Amplitudes and the Double Copy in Topologically
  Massive Theories},'' {\em JHEP} {\bf 12} (2020) 163,
  \href{http://www.arXiv.org/abs/2006.15957}{{\tt 2006.15957}}.

\bibitem{Gonzalez:2021bes}
M.~C. Gonz\'alez, A.~Momeni, and J.~Rumbutis, ``{Massive double copy in three
  spacetime dimensions},'' {\em JHEP} {\bf 08} (2021) 116,
  \href{http://www.arXiv.org/abs/2107.00611}{{\tt 2107.00611}}.

\bibitem{Burger:2021wss}
D.~J. Burger, W.~T. Emond, and N.~Moynihan, ``{Anyons and the double copy},''
  {\em JHEP} {\bf 01} (2022) 017,
  \href{http://www.arXiv.org/abs/2103.10416}{{\tt 2103.10416}}.

\bibitem{Gonzalez:2021ztm}
M.~C. Gonz\'alez, A.~Momeni, and J.~Rumbutis, ``{Massive double copy in the
  high-energy limit},'' {\em JHEP} {\bf 04} (2022) 094,
  \href{http://www.arXiv.org/abs/2112.08401}{{\tt 2112.08401}}.

\bibitem{Moynihan:2019bor}
N.~Moynihan, ``{Kerr-Newman from Minimal Coupling},'' {\em JHEP} {\bf 01}
  (2020) 014,
\href{http://www.arXiv.org/abs/1909.05217}{{\tt 1909.05217}}.
%%CITATION = ARXIV:1909.05217;%%.

\bibitem{Penrose:1987uia}
R.~Penrose and W.~Rindler, {\em {Spinors and Space-Time}}.
\newblock Cambridge Monographs on Mathematical Physics. Cambridge Univ. Press,
  Cambridge, UK, 4, 2011.

\bibitem{Stewart:1990uf}
J.~Stewart, {\em {Advanced general relativity}}.
\newblock Cambridge Monographs on Mathematical Physics. Cambridge University
  Press, 4, 1994.

\bibitem{Witten:1959zza}
L.~Witten, ``{Invariants of General Relativity and the Classification of
  Spaces},'' {\em Phys. Rev.} {\bf 113} (1959) 357--362.

\bibitem{Monteiro:2021ztt}
R.~Monteiro, S.~Nagy, D.~O'Connell, D.~Peinador~Veiga, and M.~Sergola, ``{NS-NS
  spacetimes from amplitudes},'' {\em JHEP} {\bf 06} (2022) 021,
  \href{http://www.arXiv.org/abs/2112.08336}{{\tt 2112.08336}}.

\bibitem{Newman:1965tw}
E.~T. Newman and A.~I. Janis, ``{Note on the Kerr spinning particle metric},''
  {\em J. Math. Phys.} {\bf 6} (1965) 915--917.

\bibitem{deUrreta:2015nla}
E.~J.~G. de~Urreta and M.~Socolovsky, ``{Extended Newman-Janis algorithm for
  rotating and Kerr-Newman de Sitter and anti de Sitter metrics},''
  \href{http://www.arXiv.org/abs/1504.01728}{{\tt 1504.01728}}.

\end{thebibliography}\endgroup
\end{document}